\documentclass{aastex631}

\usepackage{bm}
\usepackage{subcaption}
\usepackage[T1]{fontenc}
\usepackage{booktabs}
\usepackage{threeparttable}
\usepackage{multirow}
\usepackage{amsmath}
\usepackage[utf8]{inputenc}
\usepackage{makecell}
\usepackage{hyperref}
\usepackage{numprint}
\usepackage[ruled,vlined]{algorithm2e}
\usepackage{tabularx}

\begin{document}

 \title{A Kinematic Constraint on Pedestrian Walking: Power-law Scaling between Critical Angular Velocity and Speed}
 
\author{Jinghui Wang}

\affiliation{School of Safety Science and Emergency Management\\
Wuhan University of Technology\\
Wuhan, China}

\author{Wei Lv}
\affiliation{School of Safety Science and Emergency Management\\
Wuhan University of Technology\\
Wuhan, China}
\affiliation{China Research Center for Emergency Management\\
Wuhan University of Technology\\
Wuhan, China}

\author{Chao Li}

\affiliation{School of Architecture and Urban Planning\\
Chongqing University\\
Chongqing, China}
\affiliation{Key Laboratory of New Technology for Construction of Cities in Mountain Area\\
Chongqing University\\
Chongqing, China}

\author{Yufei Li}

\affiliation{School of Automation\\
Central South University\\
Changsha, China}

\begin{abstract}
This paper presents a statistical analysis of speed and angular velocity obtained from pedestrian experiments across nine distinct datasets. Experimental scenarios included crossing motion, unidirectional/bidirectional flows, bidirectional/four-directional crossing flows, pedestrian-vehicle interactions, unidirectional flow in a circular corridor, and circle antipode configurations. We applied filtering methods to reduce noise and analyzed the data at different sampling frequencies. The results reveal a universal power-law scaling between critical angular velocity and speed, with a scaling exponent of approximately -0.8. This relationship defines a bounded region in the speed–angular velocity phase space, suggesting a kinematic constraint on pedestrian motion.
\end{abstract}


\section{Introduction} 
\label{section1}

From pedestrian modelling to crowd management, quantitative metrics are essential for evaluating pedestrian flow. Fundamental diagrams, derived from vehicular traffic theory \citep{greenshields}, offer a basic yet widely used theoretical basis. Direct observation of flow, speed, and density provided critical insights into pedestrian dynamics. Research on pedestrian fundamental diagrams, based on empirical data, has identified two key relationships: the speed-density and flow-density relationships, both typically described by empirical functions.

A essential distinction in pedestrian dynamics arises from its spatial differences compared with vehicular traffic. Conventional traffic flow studies examine one-dimensional continuous flow, which is traditionally modeled as a linear dynamical system \citep{cordes2023single}. Even incorporating lane-changing behavior, vehicular systems can be approximated as 1.5-dimensional dynamical systems (constrained two-dimensional motion). In contrast, pedestrian motion typically occurs in open spaces and exhibits inherently two-dimensional dynamics. Unlike vehicles, pedestrians participate in traffic flows without relying on carriers, resulting in highly flexible motion patterns. The heterogeneous nature of pedestrian motion manifests itself in distinct fundamental diagram characteristics across different scenarios \citep{hoogendoorn2011fundamental}. Furthermore, the dimensional mismatch between vehicular and pedestrian motion often limits the applicability of conventional evaluation metrics \citep{saberi2014exploring}. As a consequence, the distinction between areawide and generalized evaluation has led to the development of micro and macro fundamental diagrams. Despite these limitations, the fundamental diagrams remain the most widely accepted paradigm in pedestrian research. Extensive controlled experiments have yielded comprehensive empirical datasets, enabling systematic exploration and validation \citep{seyfried2005fundamental,zhang2012ordering,cao2017fundamental}.

Directional adjustments constitute a prevalent feature of pedestrian locomotion. This characteristic has been explored through empirical observations \citep{hicheur2005velocity,parisi2016experimental}. Pedestrian acceleration is typically associated with a decrease in angular variation. However, the intrinsic correlation between angular velocity and speed during pedestrian motion has not been systematically investigated. A representative case occurs when pedestrians reverse direction, often exhibiting unicycle-like trajectories \citep{farina2017walking}. Alternatively, pedestrians may decelerate to standstill before turning, highlighting the inherent trade-off between speed and angular variation. This reciprocal relationship governs normal pedestrian navigation \citep{dias2013experimental,huber2014adjustments,elliott2012pedestrian} and manifests even in competitive sports \citep{dos2018effect}, underscoring its universal applicability.

Despite being a prominent feature of pedestrian movement, directional variation has long been overlooked. In this study, we investigated the relationship between curvature and speed through empirical analysis of pedestrian trajectories from 9 independent datasets. Our statistical results have revealed significant findings. The paper is organized as follows: Section \ref{section2} describes the experimental scenarios and configurations, where pedestrian trajectories were reconstructed from preprocessed data. Through statistical analysis of trajectory data, Section \ref{section3} systematically examines the speed-curvature relationship via scatter plots. These empirical results demonstrate a distinct power-law constraint governing pedestrian dynamics. Finally, the results of this paper are discussed in Section \ref{section4}.

\section{Experimental setup} \label{section2}

\subsection{Datasets} \label{subsection2.1}

\begin{figure}[ht!]
\centering
\includegraphics[scale=0.5]{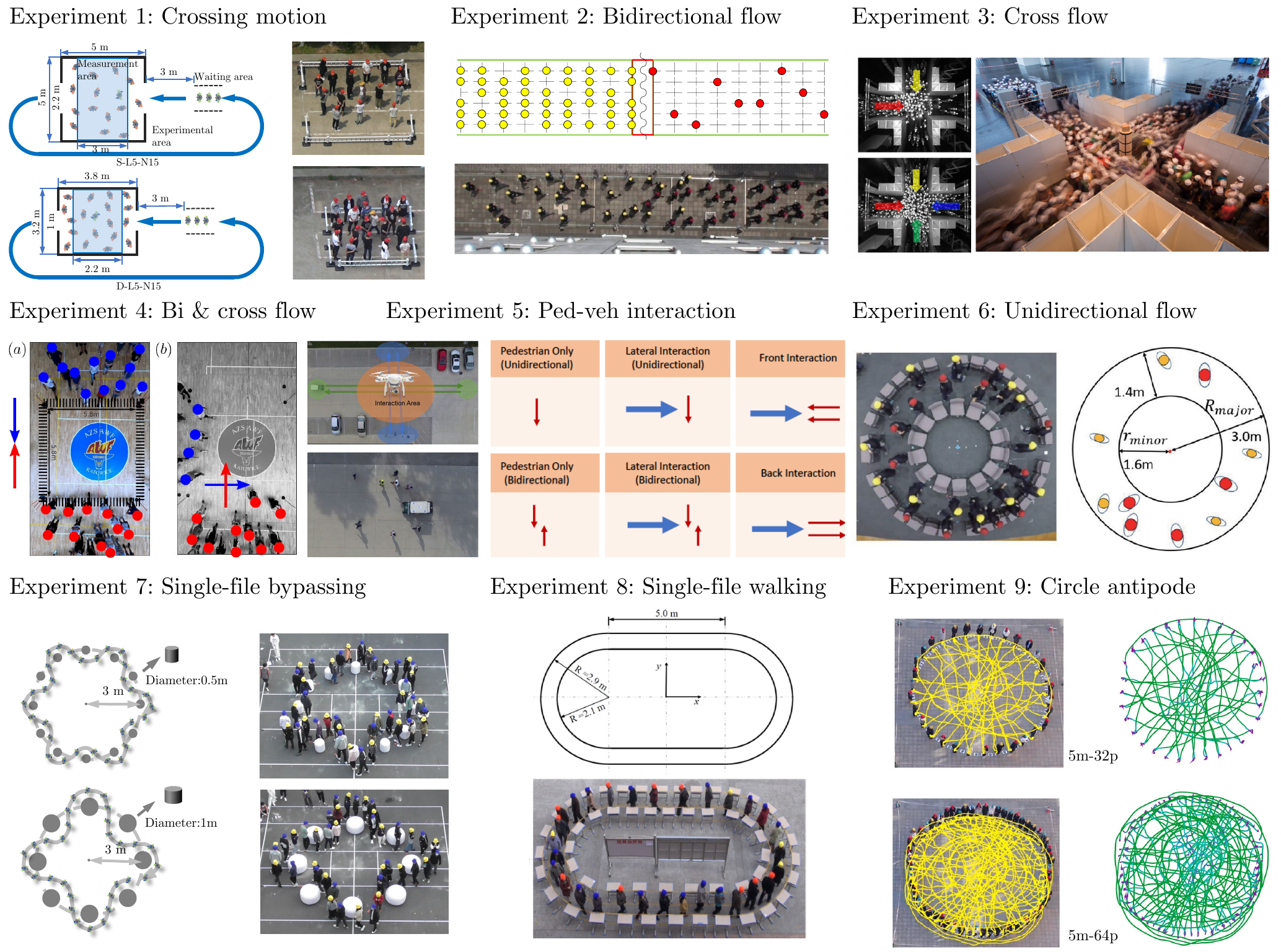}
\caption{Graphic summary of the experiments.}
\label{fig1}
\end{figure}

Our data were collected from 9 controlled pedestrian experiments. Among these, two sets of experiments (experiment 1 and experiment 7) were conducted and extracted by the authors, while the majority of the remaining experimental data were obtained from public datasets. Detailed information for each experiment is provided in Table \ref{table1}, and a graphical summary is presented in Fig. \ref{fig1}.

\begin{table}[htbp]
\centering
\caption{Summary of experimental datasets}
{\scriptsize
\begin{tabular}{ccccccc}
\toprule
Datasets & Scenarios & Participants & Data Acquisition & Frame Rate (Hz) & Sampling Interval (s) & Preprocessing \\
\midrule
Experiment 1 & Crossing motion & 50 & Camera (manual+DLT) & 25 & 0.04, 0.4 & Mean filtering \\
Experiment 2 & Bidirectional flow & 54 & Camera (PeTrack) & 30 & 1/30, 1/3 & Adaptive filtering \\
Experiment 3 & Cross flow & $ \approx 2000 $ & Camera (PeTrack) & 25 (A), 16 (D) & 0.04, 0.4; 0.0625, 0.625 & Adaptive filtering \\
Experiment 4 & Bi-flow \& crossflow & 73 & Camera (Matlab script) & 10 & 0.1, 1 & Adaptive filtering \\
Experiment 5 & Ped-veh interaction & $ \approx 15 $ & Drone (KNN algorithm) & 29.97 & 1/29.97, 1/2.997 & Kalman filtering \\
Experiment 6 & Unidirectional flow & 32 & Camera (PeTrack) & 30 & 1/30, 1/3 & Adaptive filtering \\
Experiment 7 & Single-file bypassing & 60 & Camera (PeTrack) & 25 & 0.04, 0.4 & Adaptive filtering \\
Experiment 8 & Single-file walking & 127 & Camera (PeTrack) & 25 & 0.04, 0.4 & Adaptive filtering \\
Experiment 9 & Circle antipode & 64 & Camera (PeTrack) & 25 & 0.04, 0.4 & None (Raw data) \\
\bottomrule
\end{tabular}
}
\label{table1}
\end{table}

The raw trajectory data contained errors originating from two primary sources. The first source was associated with the data extraction process, in which spatial coordinates were obtained from video data, either manually or through automatic image recognition methods, will introducing unavoidable inaccuracies that resulted in jagged trajectories. The second source originated from the common practice of using the head position to represent the pedestrian’s 2D coordinates during extraction. However, during motion, natural head oscillations acted as a form of locomotion compensation, preventing the head position from fully reflecting the motion of the torso. To address these issues, preprocessing was performed on the raw data. In Experiment 1, mean filtering was applied, with the coordinate processing defined by Eq. \ref{1}, where $k$ denotes the frame index. 

\begin{align}\label{1}
\tilde{\mathbf{x}}_n = \frac{1}{2k+1} \sum_{i=-k}^{k} \mathbf{x}_{n+i}, \quad k=5
\end{align}

The trajectory data from the Experiment 5 had already been filtered using the method of Kalman filter.  No filtering was applied to the data from Experiment 9 for comparison, and the remaining experiments were denoised using an adaptive filtering method, specifically the Normalized Least Mean Squares (NLMS) algorithm \citep{mathworks_lms_nlms}.

\subsection{Experiment descriptions} \label{subsection2.1}

• \textbf{Experiment 1: pedestrian crossing motion within static and dynamic crowds}

Experiment 1 aims to investigate pedestrian crossing behaviors in static and dynamic crowds. The experiment consists of two parts: static trials and dynamic crowd trials. In each trial, a predetermined number of participants either stood stationary or walked freely within the experimental area. At the beginning of each repetition, crossing pedestrians gathered in a designated waiting area. Upon the start signal, they sequentially crossed through the crowd. Each group completed the experiment after a fixed number of repetitions. The study aimed to analyze how pedestrians interact with crowds by adjusting their speed or direction to avoid collisions \citep{nicolas2019mechanical,kleinmeier2020agent}.

\begin{figure}[ht!]
\centering
\includegraphics[scale=0.3]{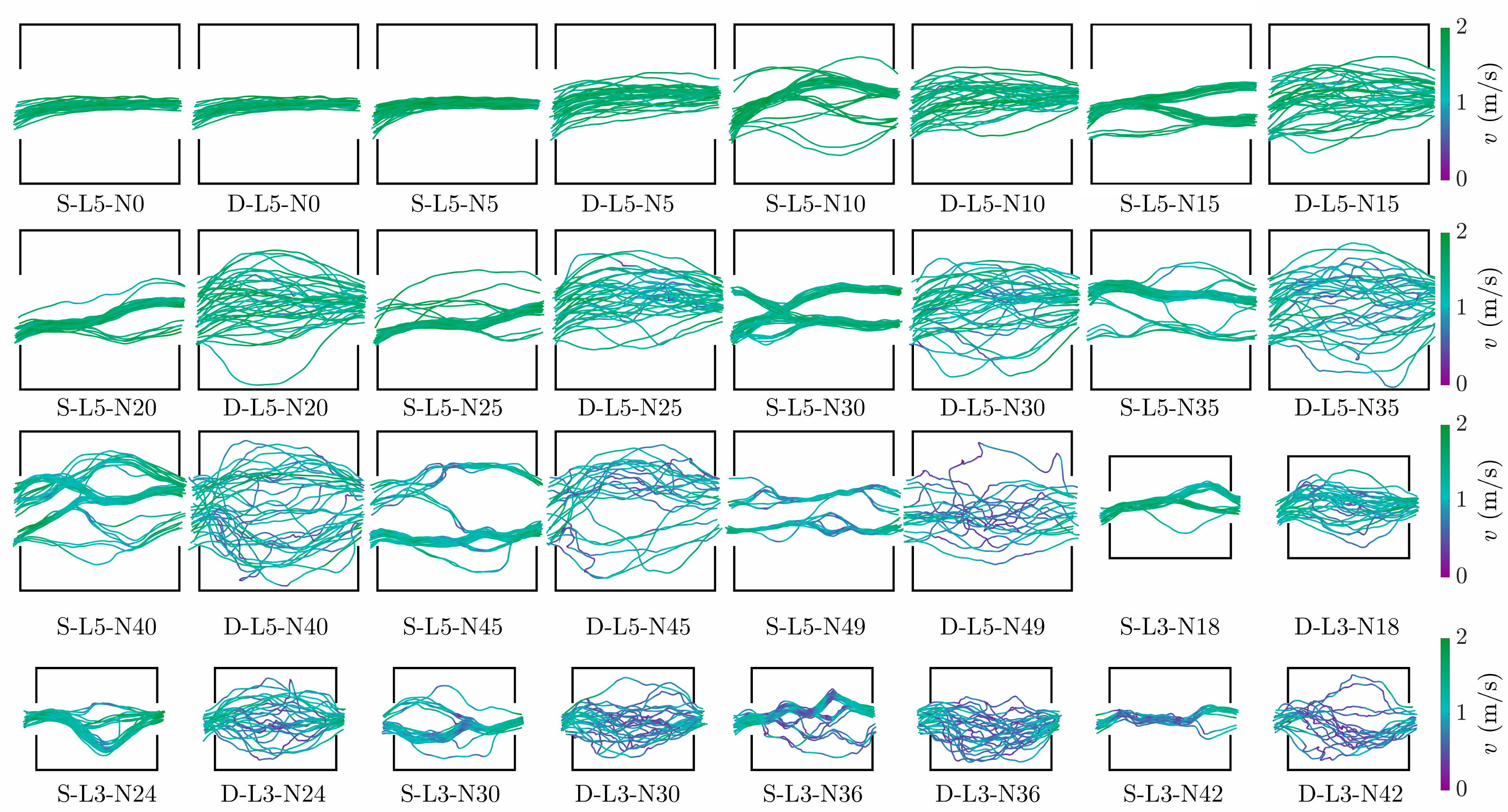}
\caption{Pedestrian trajectories from Experiment 1, with the corresponding indices labeled below.}
\label{fig2}
\end{figure}

For detailed procedures, please refer to \citep{wang2023exploring}. Figs. \ref{fig1} and \ref{fig2} show an experiment snapshot and preprocessed trajectories, respectively. Data are available at: \url{https://ped.fz-juelich.de/database/doku.php?id=crossing_static_dynamic}.

\begin{table}[]
\nolinenumbers
\caption{Setup for Experiment 1.}
\centering
\begin{tabular}{*{6}{c}} 
    \toprule
    Index & Crowd context & Configuration (m\textsuperscript{2}) & 
    \makecell{No. of\\ Participants} & 
    \makecell{Global density\\ (ped/m\textsuperscript{2})} & 
    \makecell{Repetitions} \\
    \midrule
    S(D)-L5-N0  & Static (Dynamic) & 5$\times$5        & 0  & 0    & 30      \\
    S(D)-L5-N5  & Static (Dynamic) & 5$\times$5        & 5  & 0.2  & 50 (39) \\
    S(D)-L5-N10 & Static (Dynamic) & 5$\times$5        & 10 & 0.4  & 50 (39) \\
    S(D)-L5-N15 & Static (Dynamic) & 5$\times$5        & 15 & 0.6  & 49 (39) \\
    S(D)-L5-N20 & Static (Dynamic) & 5$\times$5        & 20 & 0.8  & 50 (40) \\
    S(D)-L5-N25 & Static (Dynamic) & 5$\times$5        & 25 & 1    & 49 (38) \\
    S(D)-L5-N30 & Static (Dynamic) & 5$\times$5        & 30 & 1.2  & 50 (39) \\
    S(D)-L5-N35 & Static (Dynamic) & 5$\times$5        & 35 & 1.4  & 49 (34) \\
    S(D)-L5-N40 & Static (Dynamic) & 5$\times$5        & 40 & 1.6  & 45 (35) \\
    S(D)-L5-N45 & Static (Dynamic) & 5$\times$5        & 45 & 1.8  & 51 (23) \\
    S(D)-L5-N49 & Static (Dynamic) & 5$\times$5        & 49 & 1.96 & 24 (20) \\
    S(D)-L3-N18 & Static (Dynamic) & 3.8$\times$3.2  & 18 & 1.48 & 29 (30) \\
    S(D)-L3-N24 & Static (Dynamic) & 3.8$\times$3.2  & 24 & 1.97 & 29 (29) \\
    S(D)-L3-N30 & Static (Dynamic) & 3.8$\times$3.2  & 30 & 2.47 & 29 (29) \\
    S(D)-L3-N36 & Static (Dynamic) & 3.8$\times$3.2  & 36 & 2.96 & 29 (29) \\
    S(D)-L3-N42 & Static (Dynamic) & 3.8$\times$3.2  & 42 & 3.45 & 16 (17) \\
    \bottomrule
\end{tabular}%
\label{table2}
\end{table}

• \textbf{Experiment 2: bidirectional pedestrian flow in the presence of individuals and social groups}

Experiment 2 aimed to investigate the characteristics of bidirectional pedestrian flow and the dynamics of social groups. The experiment consisted of two parts: walking individually (individual set) and walking in pairs (group set). The geometric configuration of the experimental setup and the instructions given to participants were identical in both trails. For each set, both balanced and unbalanced flow configurations were predefined. As shown in Table \ref{table3}, each configuration was repeated four times.

\begin{figure}[ht!]
\centering
\includegraphics[scale=0.43]{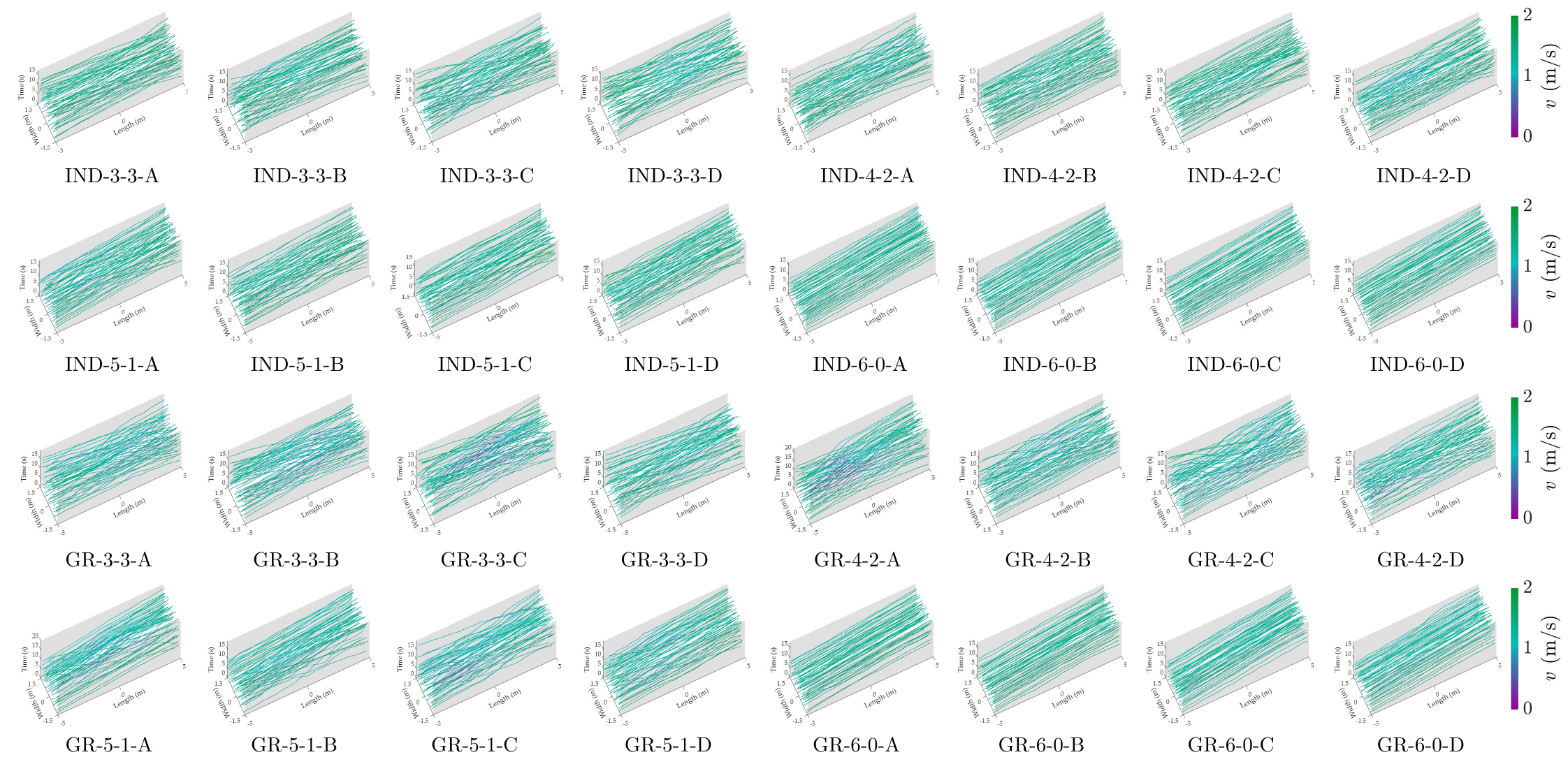}
\caption{Pedestrian trajectories from Experiment 2, with the corresponding indices labeled below.}
\label{fig3}
\end{figure}

For detailed experimental procedures, refer to \citep{feliciani2016empirical}. The preprocessed trajectories is shown in Fig. \ref{fig3}. The data for this experiment are available at: \url{https://ped.fz-juelich.de/database/doku.php?id=bidirectional_flow} (Pedestrian Dynamics Data Archive).
 
\begin{table}[htbp]
\centering
\caption{Setup for Experiment 2.}
\begin{tabular}{cccccc}
\toprule
Index & Case Type & Major flow (left) & Minor flow (right) & Flow ratio & Repetitions \\
\midrule
\multirow{4}{*}{Individual} 
  & 6/0 & 6 persons/column & 0 persons/column & 0.000 & 4 \\
  & 5/1 & 5 persons/column & 1 persons/column & 0.167 & 4 \\
  & 4/2 & 4 persons/column & 2 persons/column & 0.333 & 4 \\
  & 3/3 & 3 persons/column & 3 persons/column & 0.500 & 4 \\
\midrule
\multirow{4}{*}{Group} 
  & 6/0 & 6 persons/column & 0 persons/column & 0.000 & 4 \\
  & 5/1 & 5 persons/column & 1 persons/column & 0.167 & 4 \\
  & 4/2 & 4 persons/column & 2 persons/column & 0.333 & 4 \\
  & 3/3 & 3 persons/column & 3 persons/column & 0.500 & 4 \\
\bottomrule
\end{tabular}
\label{table3}
\end{table}

• \textbf{Experiment 3: pedestrian flow at a perpendicular crossing}

Experiment 3 was designed to investigate the characteristics of cross pedestrian flows. The study analyzed bidirectional and four-directional cross-flow experiments conducted under the BaSiGo program. Before each run, participants were instructed to move in a  "intended direction". In these cross-flow experiments, pedestrians moved toward the direction opposite their initial positions: those starting from the left (right, top, or bottom) of the crossing area proceeded to the right (left, bottom, or top), respectively. The corridor on each side of the crossing area was 5 meters in length. For both bidirectional and four-directional flow experiments, the width of the entrance bins was systematically varied. Detailed settings for all experimental runs are provided in Table \ref{table4}.

\begin{figure}[ht!]
\centering
\includegraphics[scale=0.55]{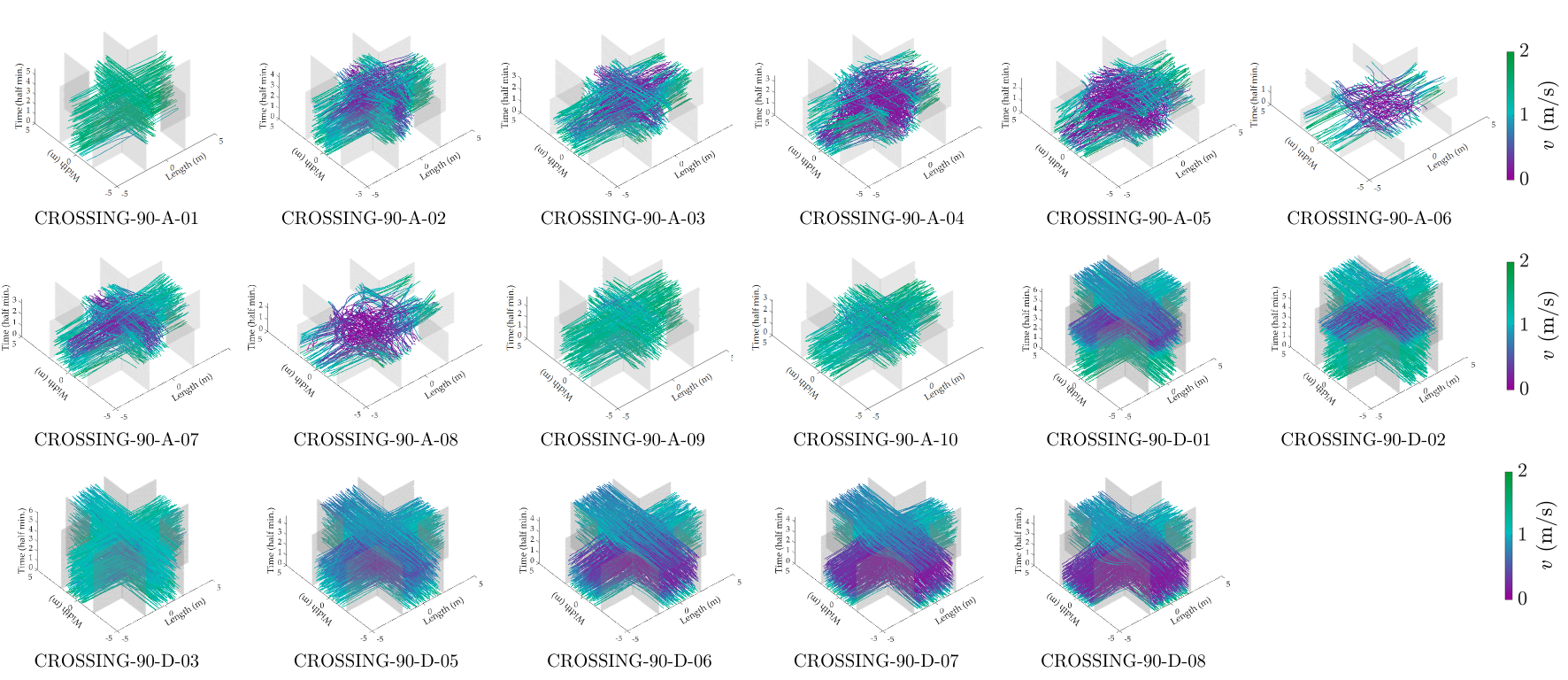}
\caption{Pedestrian trajectories from Experiment 3, with the corresponding indices labeled below.}
\label{fig4}
\end{figure}

For a comprehensive description of the experimental procedures and additional explanations, refer to \citep{cao2017fundamental}. The preprocessed trajectories is shown in Fig. \ref{fig4}. The data for this experiment are available at: \url{https://ped.fz-juelich.de/da/doku.php?id=crossing_90} (Pedestrian Dynamics Data Archive).

\begin{table}[htbp]
\centering
\caption{Setup for Experiment 3.}
\begin{tabular}{ccccc}
\toprule
Geometry & Index &  $b_{\text{in}}$ (m) &  $b_{\text{cor}}$ (m) & No. of Participants \\
\midrule
\multirow{10}{*}{Four-directional} 
& CROSS\_90\_A\_01 & 0.6 & 4.0 & 247 \\
& CROSS\_90\_A\_02 & 0.6 & 4.0 & 439 \\
& CROSS\_90\_A\_03 & 0.9 & 4.0 & 323 \\
& CROSS\_90\_A\_04 & 1.2 & 4.0 & 352 \\
& CROSS\_90\_A\_05 & 1.5 & 4.0 & 337 \\
& CROSS\_90\_A\_06 & 2.0 & 4.0 & 269 \\
& CROSS\_90\_A\_07 & 0.6 & 4.0 & 323 \\
& CROSS\_90\_A\_08 & 1.5 & 4.0 & 298 \\
& CROSS\_90\_A\_09 & 4.0 & 4.0 & 299 \\
& CROSS\_90\_A\_10 & 4.0 & 4.0 & 324 \\
\midrule
\multirow{8}{*}{Bidirectional} 
& CROSS\_90\_D\_1  & 0.6 & 4.0 & 603 \\
& CROSS\_90\_D\_2  & 0.9 & 4.0 & 604 \\
& CROSS\_90\_D\_3  & 1.2 & 4.0 & 606 \\
& CROSS\_90\_D\_5  & 1.8 & 4.0 & 600 \\
& CROSS\_90\_D\_6  & 2.4 & 4.0 & 597 \\
& CROSS\_90\_D\_7  & 3.0 & 4.0 & 604 \\
& CROSS\_90\_D\_8  & 4.0 & 4.0 & 592 \\
\bottomrule
\end{tabular}
\label{table4}
\end{table}

• \textbf{Experiment 4: lane nucleation in complex active flows}

Experiment 4 was designed to investigate the lane nucleation mechanism in active pedestrian flows. The experimental area was a square measuring 5.8 m $\times$ 5.8 m, with its corners marked by plastic poles. In all trials, participants were instructed to walk at a normal pace while maintaining a safe distance from others and avoiding collisions. Before each trial, participants gathered in an unstructured manner (without predefined lanes) behind a designated starting line and began walking toward a specified target upon the experiment organizer's signal.

Data from scenarios 1 to 3 (part of the full experimental procedure) were preprocessed and analyzed. Detailed information on all experimental runs is provided in Table \ref{table5}. The three scenarios are described as follows:

Scenario 1 (Simple head-on cross-streams): Two groups of participants started from opposite sides of the arena and were instructed to cross to the other side. This setup facilitated the emergence of archetypal lanes aligned with the dominant direction of movement.

Scenario 2 (Chirally biased cross-streams): Two groups started from adjacent sides of the area and were instructed to cross to the opposite side while following a "pass on the right" convention. This configuration resulted in the formation of tilted lanes.

Scenario 3 (Simple perpendicular cross-streams): Two groups also began from adjacent sides of the arena and were directed to cross to the other side. This scenario led to the development of diagonal crossing lanes.

\begin{figure}[ht!]
\centering
\includegraphics[scale=0.45]{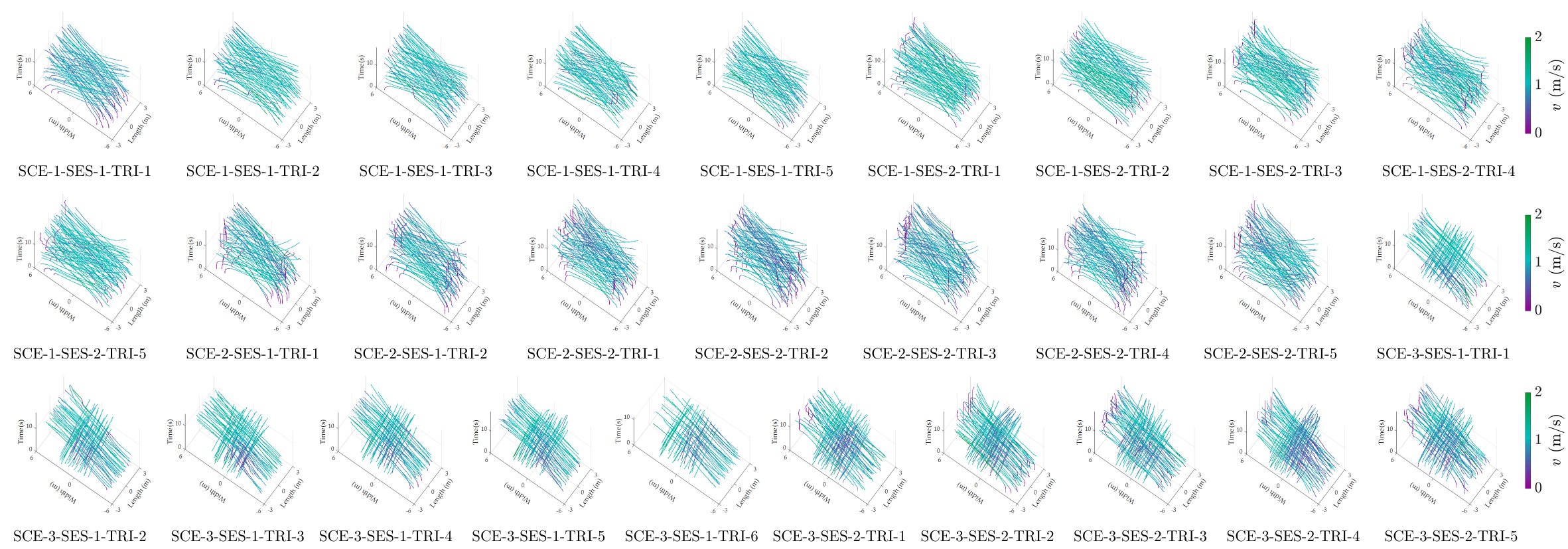}
\caption{Pedestrian trajectories from Experiment 4, with the corresponding indices labeled below.}
\label{fig5}
\end{figure}

For a detailed description of the experimental procedures and further explanations, refer to \citep{bacik2023lane}. The preprocessed trajectories is shown in Fig. \ref{fig5}. The data for this experiment were sourced from: \url{https://researchdata.bath.ac.uk/1242/} (University of Bath Research Data Archive).

\begin{table}[htbp]
\centering
\caption{Setup for Experiment 4.}
\begin{tabular}{cccc}
\toprule
Index &  Description & No. of Participants & Repetitions \\
\midrule
SCE-1-SES-1 & \multirow{2}{*}{Simple head-on cross-streams} & 60 & 5 \\
SCE-1-SES-2 & & 73 & 5 \\
\midrule
SCE-2-SES-1 & \multirow{2}{*}{Chirally biased cross-streams} & 60 & 2 \\
SCE-2-SES-2 & & 73 & 5 \\
\midrule
SCE-3-SES-1 & \multirow{2}{*}{Simple perpendicular cross-streams} & 60 & 6 \\
SCE-3-SES-2 & & 73 & 5 \\
\bottomrule
\end{tabular}
\label{table5}
\end{table}

• \textbf{Experiment 5: pedestrian-vehicle interaction}

Experiment 5 (CITR dataset) was designed to investigate vehicle-pedestrian interactions. The dataset was collected during a series of controlled experiments conducted in a parking lot near the Control and Intelligent Transportation Research Laboratory (CITR) at The Ohio State University. These experiments involved fundamental interaction scenarios between a manually driven EZ-GO golf cart and pedestrians, with the goal of analyzing pedestrian movement behavior under vehicular influence. A DJI Phantom 3 SE drone recorded the entire process. Participants were instructed to walk from a designated starting area to a specified destination area while the golf cart was operated by a human driver. All trajectory data were denoised using Kalman filtering. Detailed information about the experimental runs is provided in Table \ref{table6}.

\begin{figure}[ht!]
\centering
\includegraphics[scale=0.48]{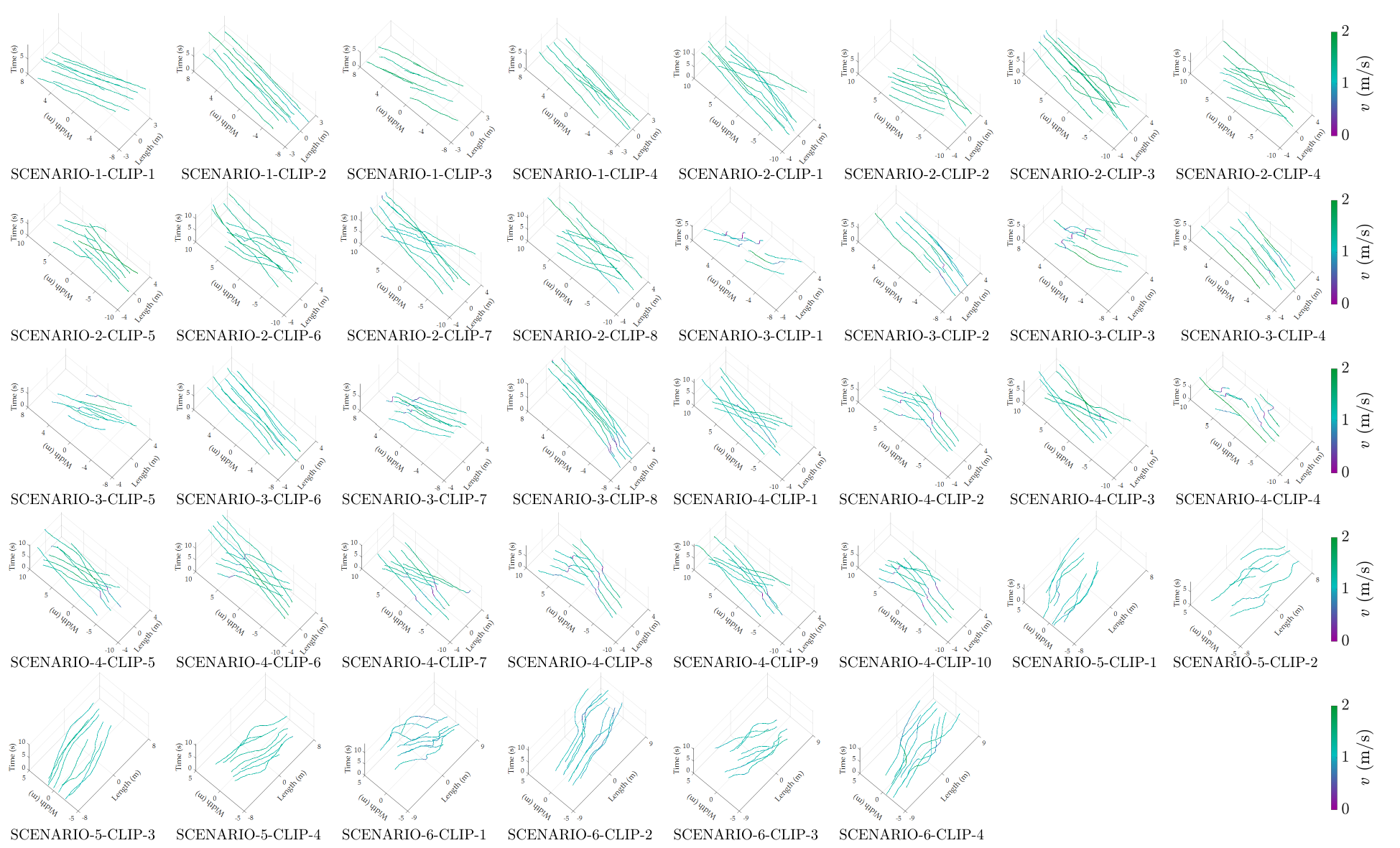}
\caption{Pedestrian trajectories from Experiment 5, with the corresponding indices labeled below.}
\label{fig6}
\end{figure}

For a comprehensive description of the experimental procedures, refer to \citep{yang2019top}. The preprocessed trajectories is shown in Fig. \ref{fig6}. The data is available at: \url{https://github.com/dongfang-steven-yang/vci-dataset-citr}. All original data were denoised using the Kalman filter \citep{yang2019top}.

\begin{table}[htbp]
\centering
\caption{Setup for Experiment 5.}
\begin{tabular}{cccc}
\toprule
Index & Scenarios \& Description & Flow Type & No. of Clips \\
\midrule
SCENARIO-1 & Pedestrian only           & Unidirectional & 4  \\
SCENARIO-2 & Pedestrian only           & Unidirectional & 8  \\
SCENARIO-3 & Lateral interaction       & Bidirectional  & 8  \\
SCENARIO-4 & Lateral interaction       & Bidirectional  & 10 \\
SCENARIO-5 & Front interaction         & Unidirectional & 4  \\
SCENARIO-6 & Back interaction          & Unidirectional & 4  \\
\bottomrule
\end{tabular}
\label{table6}
\end{table}

• \textbf{Experiment 6: unidirectional pedestrian flow at different speeds}

Experiment 6 investigated pedestrian movement dynamics in crowds with varying walking speeds. The study focused on three pedestrian types: slow walkers, normal walkers, and fast walkers. This design aimed to analyze how speed heterogeneity influences individual behavior and overall flow characteristics. Each experimental group was further divided into eight cases based on incremental density settings. In all cases, pedestrians walked unidirectionally along a circular path.

\begin{figure}[ht!]
\centering
\includegraphics[scale=0.45]{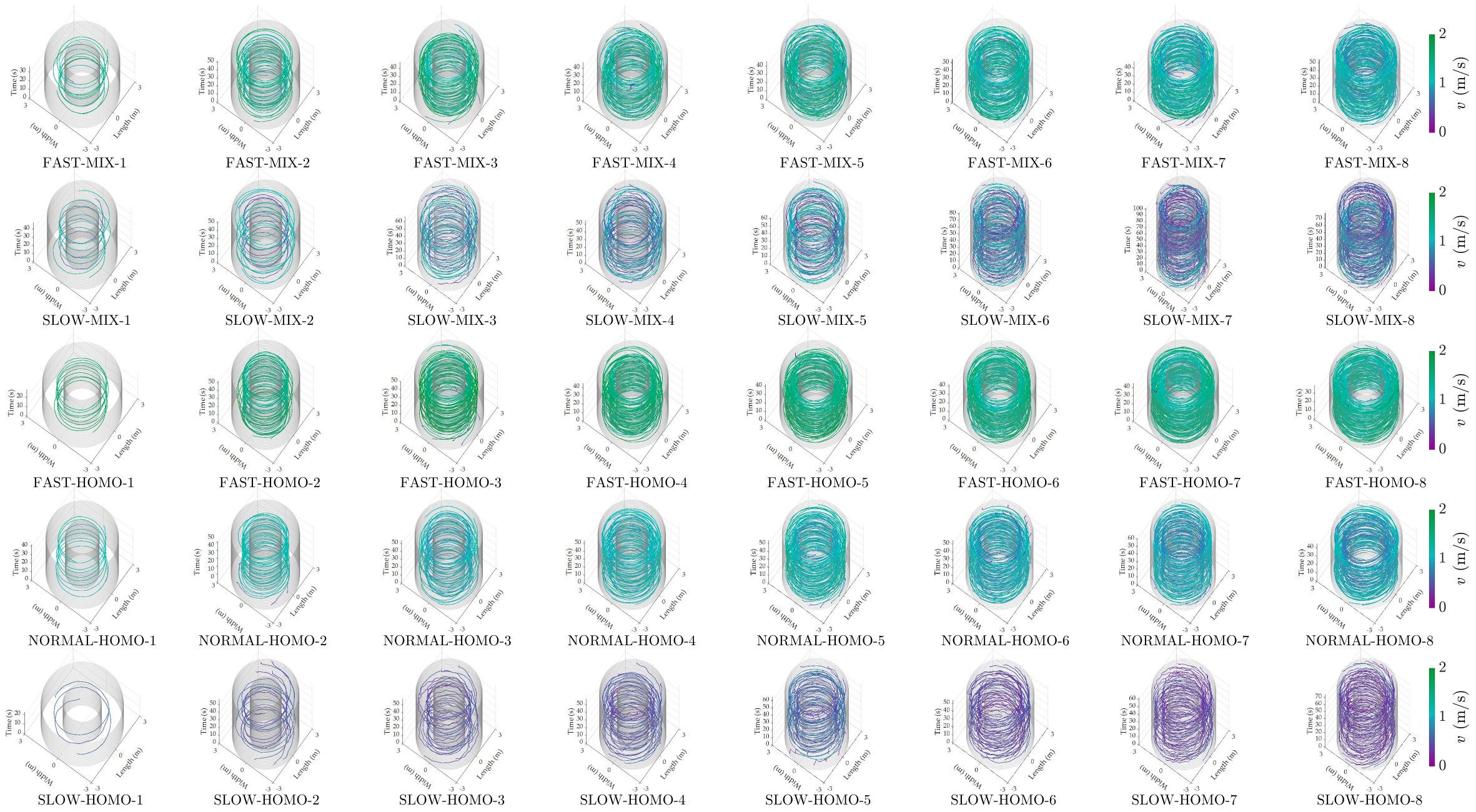}
\caption{Pedestrian trajectories from Experiment 6, with the corresponding indices labeled below.}
\label{fig7}
\end{figure}

Detailed information for all experimental runs is provided in Table \ref{table7}. For a comprehensive description of the experimental procedures, refer to \citep{fujita2019traffic}. The preprocessed trajectories is shown in Fig. \ref{fig7}.  The data for this experiment are available at: \url{https://ped.fz-juelich.de/database/doku.php?id=traffic_flow_at_different_speeds}(Pedestrian Dynamics Data Archive).

\begin{table}[htbp]
\centering
\caption{Setup for Experiment 6.}
\begin{tabular}{cccc}
\toprule
Index & Mixed elements & Global density (ped/m\textsuperscript{2}) & No. of cases \\
\midrule
SLOW-MIX       & Fast and normal         & 0.1--1.5  (increments of 0.2) & 8 \\
FAST-MIX       & Slow and normal         & 0.1--1.5  (increments of 0.2) & 8 \\
SLOW-HOMO      & Fast walkers only       & 0.1--1.5  (increments of 0.2) & 8 \\
NORMAL-HOMO    & Normal walkers only     & 0.1--1.5  (increments of 0.2) & 8 \\
FAST-HOMO      & Slow walkers only       & 0.1--1.5  (increments of 0.2) & 8 \\
\bottomrule
\end{tabular}
\label{table7}
\end{table}

• \textbf{Experiment 7: continuous obstacle-avoidance dynamics of single-file pedestrians}

Experiment 7 was designed to investigate the motion characteristics of single-file pedestrians during continuous detour maneuvers. A total of 60 participants (30 males and 30 females, aged 21–26) engaged in the experiment. The experiment was conducted in an 8 m $\times$ 8 m measurement area, where 8 (or 12) cylindrical obstacles with diameters of 1 m (or 0.5 m) and heights of approximately 0.5 m were uniformly arranged in a circular formation. Before the experiment, predefined movement directions were marked on the ground to guide participants. In mixed-gender trials, male and female participants were arranged alternately. Each trial lasted for over two minutes, with a 40-second segment (1000 frames) of pedestrian coordinate data extracted for analysis.

\begin{figure}[ht!]
\centering
\includegraphics[scale=0.55]{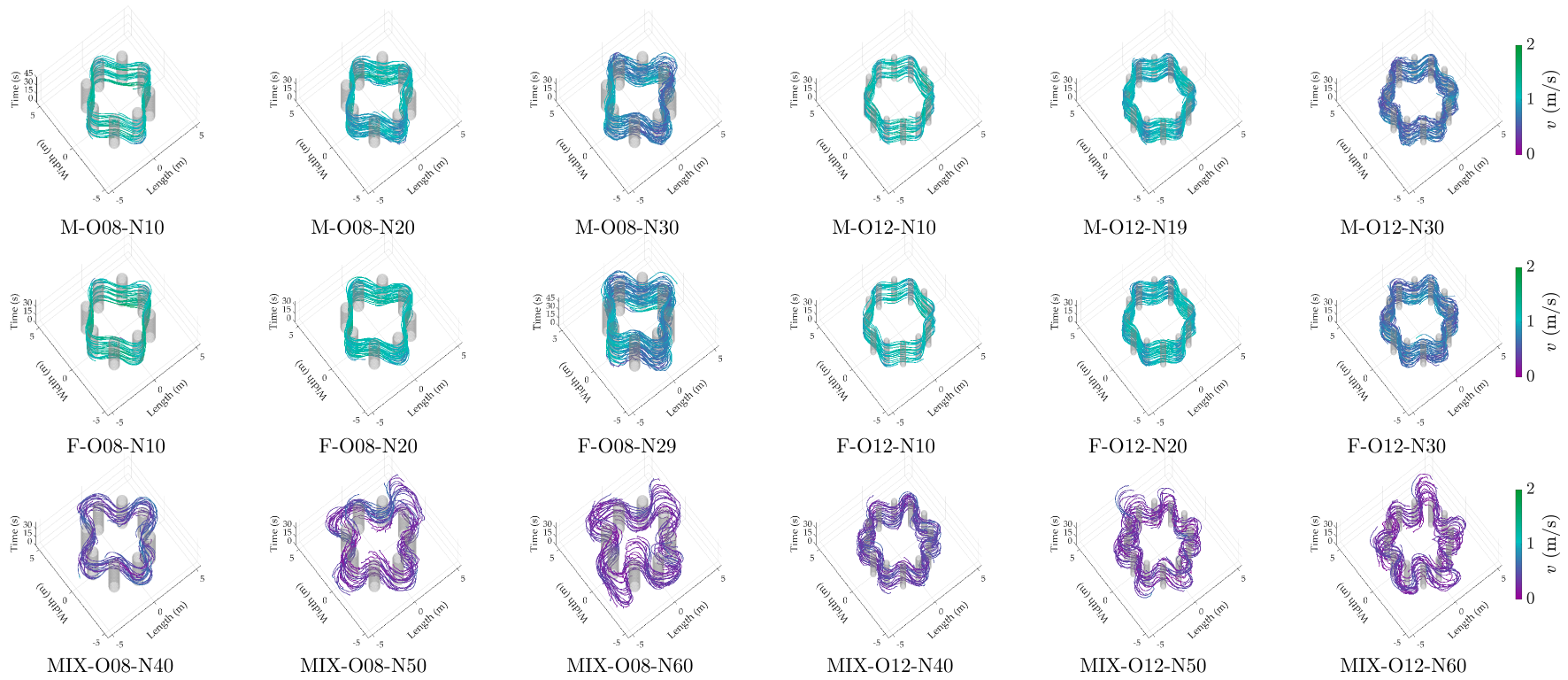}
\caption{Pedestrian trajectories from Experiment 7, with the corresponding indices labeled below.}
\label{fig8}
\end{figure}

Detailed information for all experimental runs is provided in Table \ref{table8}, and the preprocessed trajectories is shown in Fig. \ref{fig8}. The experimental data are available at:
\url{https://drive.google.com/drive/folders/1t2CvkysYB5CKJ-vqNi6Z45lbO2V1g6Tl}.

\begin{table}[htbp]
\centering
\caption{Setup for Experiment 7.}
\begin{tabular}{ccccc|ccccc}
\toprule
Index & Obstacles & Male & Female & Total & Index & Obstacles & Male & Female & Total \\
\midrule
M-O08-N10   & 8  & 10 & -- & 10 & M-O12-N10   & 12 & 10 & -- & 10 \\
M-O08-N20   & 8  & 20 & -- & 20 & M-O12-N19   & 12 & 19 & -- & 20 \\
M-O08-N30   & 8  & 30 & -- & 30 & M-O12-N30   & 12 & 30 & -- & 30 \\
F-O08-N10   & 8  & -- & 10 & 10 & F-O12-N10   & 12 & -- & 10 & 10 \\
F-O08-N20   & 8  & -- & 20 & 20 & F-O12-N20   & 12 & -- & 20 & 20 \\
F-O08-N29   & 8  & -- & 29 & 29 & F-O12-N30   & 12 & -- & 30 & 30 \\
MIX-O08-N40 & 8  & 20 & 20 & 40 & MIX-O12-N40 & 12 & 20 & 20 & 40 \\
MIX-O08-N50 & 8  & 25 & 25 & 50 & MIX-O12-N50 & 12 & 25 & 25 & 50 \\
MIX-O08-N60 & 8  & 30 & 30 & 60 & MIX-O12-N60 & 12 & 30 & 30 & 60 \\
\bottomrule
\end{tabular}
\label{table8}
\end{table}

• \textbf{Experiment 8: influence of age to single-file motion}

Experiment 8 was conducted to investigate the movement dynamics of single-file pedestrians across different age groups. The participants consisted of 80 young students (aged 16–18 years, mean age 17) from the school and 47 older adults (aged 45–73 years, mean age of 52). In the mixed-age condition, students and older adults were arranged in the corridor alternately. To ensure sufficient data collection, participants were instructed to walk for at least three minutes without overtaking.

\begin{figure}[ht!]
\centering
\includegraphics[scale=0.58]{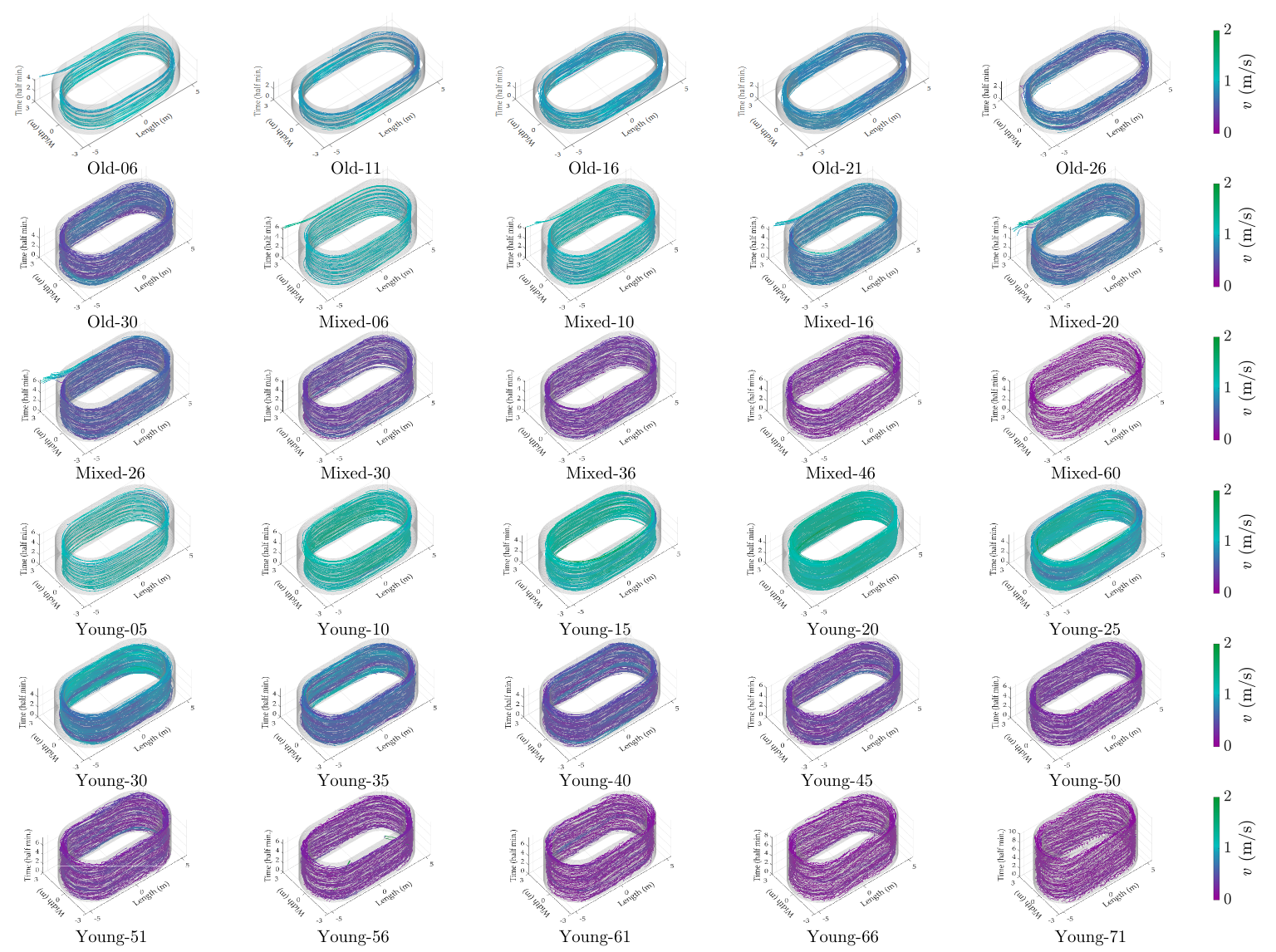}
\caption{Pedestrian trajectories from Experiment 8, with the corresponding indices labeled below.}
\label{fig9}
\end{figure}

Detailed information on all experimental runs is provided in Table \ref{table9}. For a comprehensive description of the experimental procedures and additional explanations, refer to \citep{cao2016pedestrian}. The preprocessed trajectories is shown in Fig. \ref{fig9}. The data for this experiment are available at: \url{https://ped.fz-juelich.de/da/doku.php?id=start#data_section}(Pedestrian Dynamics Data Archive).

\begin{table*}[htbp]
\centering
\caption{Setup for Experiment 8.}
\label{table9}
\begin{tabular}{ccccc|ccccc}
\toprule
 Index & Young & Old & Total & Global density (ped/m) & Index & Young & Old & Total & Global density (ped/m) \\
\midrule
 Young-05 & 5  & - & 5 & 0.19 & Old-06   & -  & 6 & 6  & 0.23 \\
 Young-10 & 10 & - & 10  & 0.39 & Old-11   & -  & 11 & 11 & 0.43 \\
 Young-15 & 15 & - & 15  & 0.58 & Old-16   & -  & 16 & 16 & 0.62 \\
 Young-20 & 20 & - & 20  & 0.78 & Old-21   & -  & 21 & 21 & 0.82 \\
 Young-25 & 25 & - & 25  & 0.97 & Old-26   & -  & 26 & 26 & 1.01 \\
 Young-30 & 30 & - & 30  & 1.17 & Old-30   & -  & 30 & 30 & 1.17 \\
 Young-35 & 35 & - & 35  & 1.36 & Mixed-06 & 3  & 3 & 6  & 0.23 \\
 Young-40 & 40 & - & 40  & 1.56 & Mixed-10 & 5  & 5 & 10  & 0.39 \\
 Young-45 & 45 & - & 45  & 1.75 & Mixed-16 & 8  & 8 & 16  & 0.62 \\
 Young-50 & 50 & - & 50  & 1.94 & Mixed-20 & 10 & 10 & 20 & 0.78 \\
 Young-51 & 51 & - & 51  & 1.98 & Mixed-26 & 13 & 13 & 26 & 1.01 \\
 Young-56 & 56 & - & 56  & 2.18 & Mixed-30 & 15 & 15 & 30 & 1.17 \\
 Young-61 & 61 & - & 61  & 2.37 & Mixed-36 & 18 & 18 & 36 & 1.40 \\
 Young-66 & 66 & - & 66  & 2.57 & Mixed-46 & 23 & 23 & 46 & 1.79 \\
 Young-71 & 71 & - & 71  & 2.76 & Mixed-60 & 30 & 30 & 60 & 2.33 \\
\bottomrule
\end{tabular}
\end{table*}

• \textbf{Experiment 9: circle antipode}

Experiment 9 was conducted to investigate collision avoidance dynamics in chaotic crowds. The experiments were performed in two circular arenas with radii of 5 m and 10 m. In each experimental group, participants (8, 16, 32, or 64) were evenly distributed along the circular boundary, with their destinations set as the antipodal points on the circle. These destinations were marked by pre-established ground markers. Each experimental condition was repeated four times.

\begin{figure}[ht!]
\centering
\includegraphics[scale=0.45]{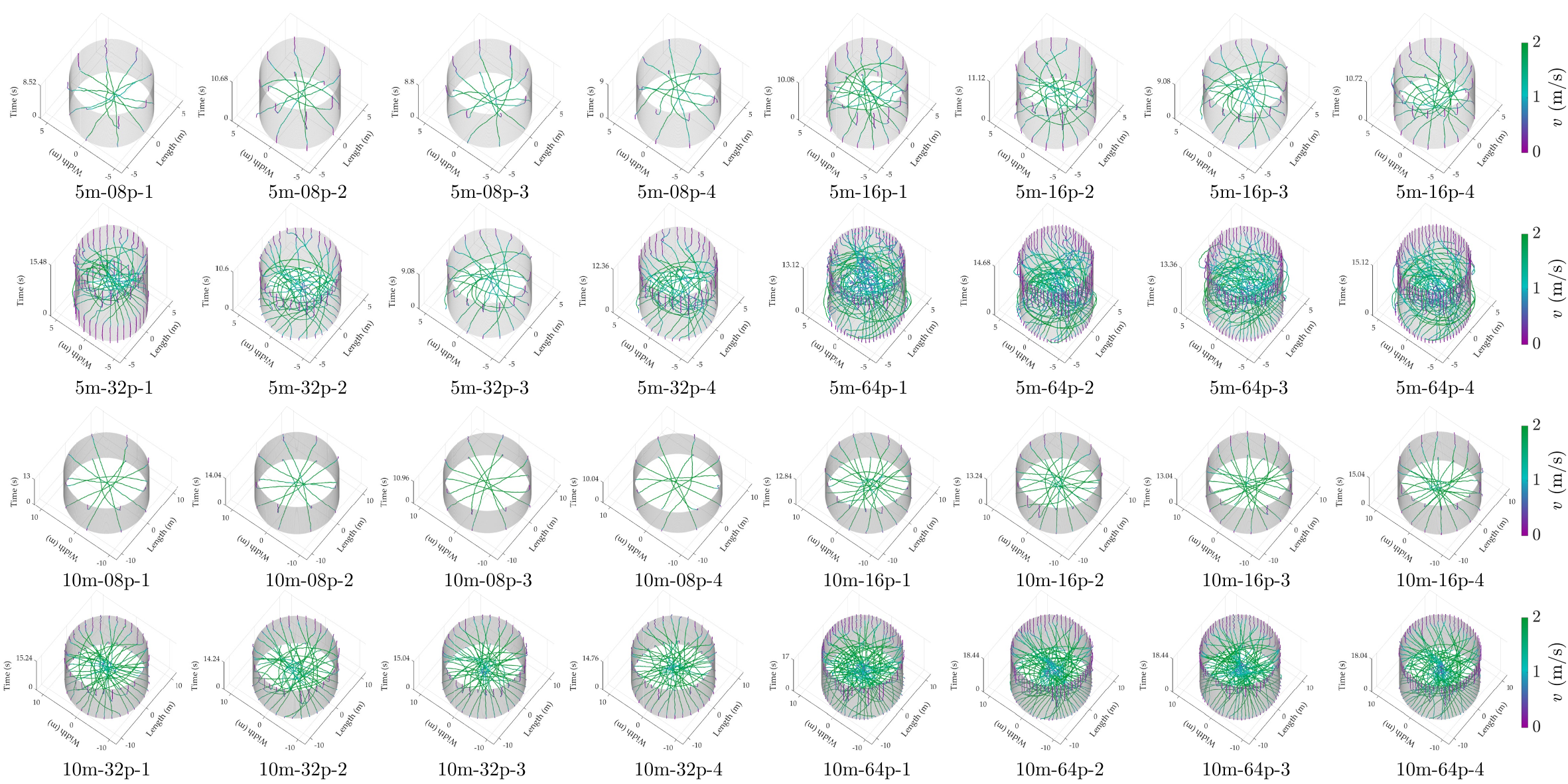}
\caption{Pedestrian trajectories from Experiment 9, with the corresponding indices labeled below.}
\label{fig10}
\end{figure}

Detailed information for all experimental runs is provided in Table \ref{table10}. For a complete description of the experimental procedures and additional explanations, refer to \citep{xiao2019investigation}. The preprocessed trajectories is shown in Fig. \ref{fig10}

\begin{table}[htbp]
\centering
\caption{Setup for Experiment 9.}
\begin{tabular}{cccc|cccc}
\toprule
Index & Radius & No. of participants & Repetitions & Index & Radius & No. of participants & Repetitions \\
\midrule
5m-08p & 5  & 8  & 4 & 10m-08p & 10 & 8  & 4 \\
5m-16p & 5  & 16 & 4 & 10m-16p & 10 & 16 & 4 \\
5m-32p & 5  & 32 & 4 & 10m-32p & 10 & 32 & 4 \\
5m-64p & 5  & 64 & 4 & 10m-64p & 10 & 64 & 4 \\
\bottomrule
\end{tabular}
\label{table10}
\end{table}

\section{Numerical statistics} \label{section3}

\subsection{Sampling configuration} \label{subsection3.1}

Curvature is defined as the rate of change of the tangent angle ($\theta$) with respect to arc length ($\ell$). An approximate method is employed for curvature calculation from discrete data. In this paper, the clockwise direction is defined as the positive orientation and the calculation procedure is detailed in Eq. \ref{2}.

\begin{align}\label{2}
\kappa &= \frac{\mathrm{d} \theta}{\mathrm{d} \ell} \simeq \frac{\alpha}{s} = \frac{\omega}{\upsilon}
\end{align}

Here, \(\alpha\) denotes the angular variation within the sampling interval, \(s\) represents the displacement during the same interval, \(\omega\) is the angular velocity, and \(v\) is the speed.

\begin{figure}[ht!]
\centering
\includegraphics[scale=0.5]{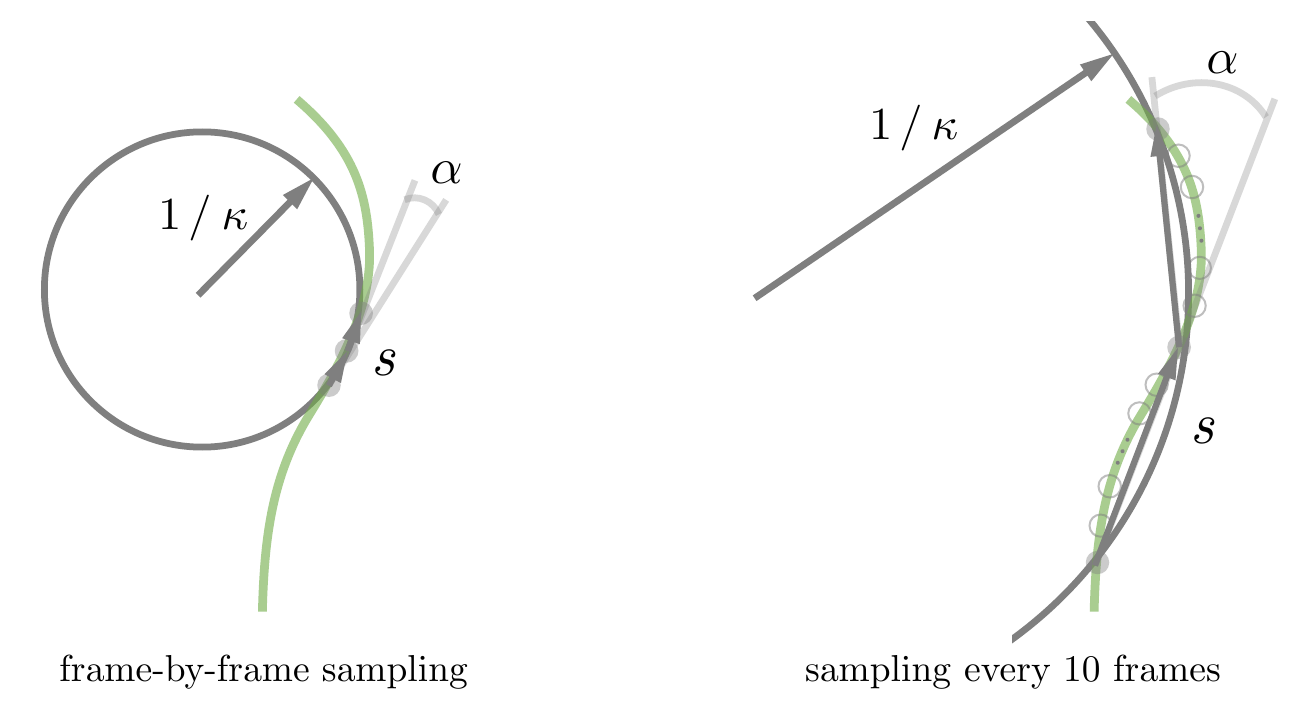}
\caption{Illustration of the curvature calculation.}
\label{fig11}
\end{figure}

Due to variations in the data acquisition setup across experiments, the time intervals per frame varied from 1/30 s (minimum) to 1/10 s (maximum). Two temporal resolutions were adopted in each experiment: frame-by-frame sampling and decimated sampling every 10 frames, as illustrated in Fig. \ref{fig11}. The specific sampling intervals are detailed in Table \ref{table1}.

\subsection{Results} \label{subsection3.2}

\begin{figure}[ht!]
\centering
\includegraphics[scale=0.66]{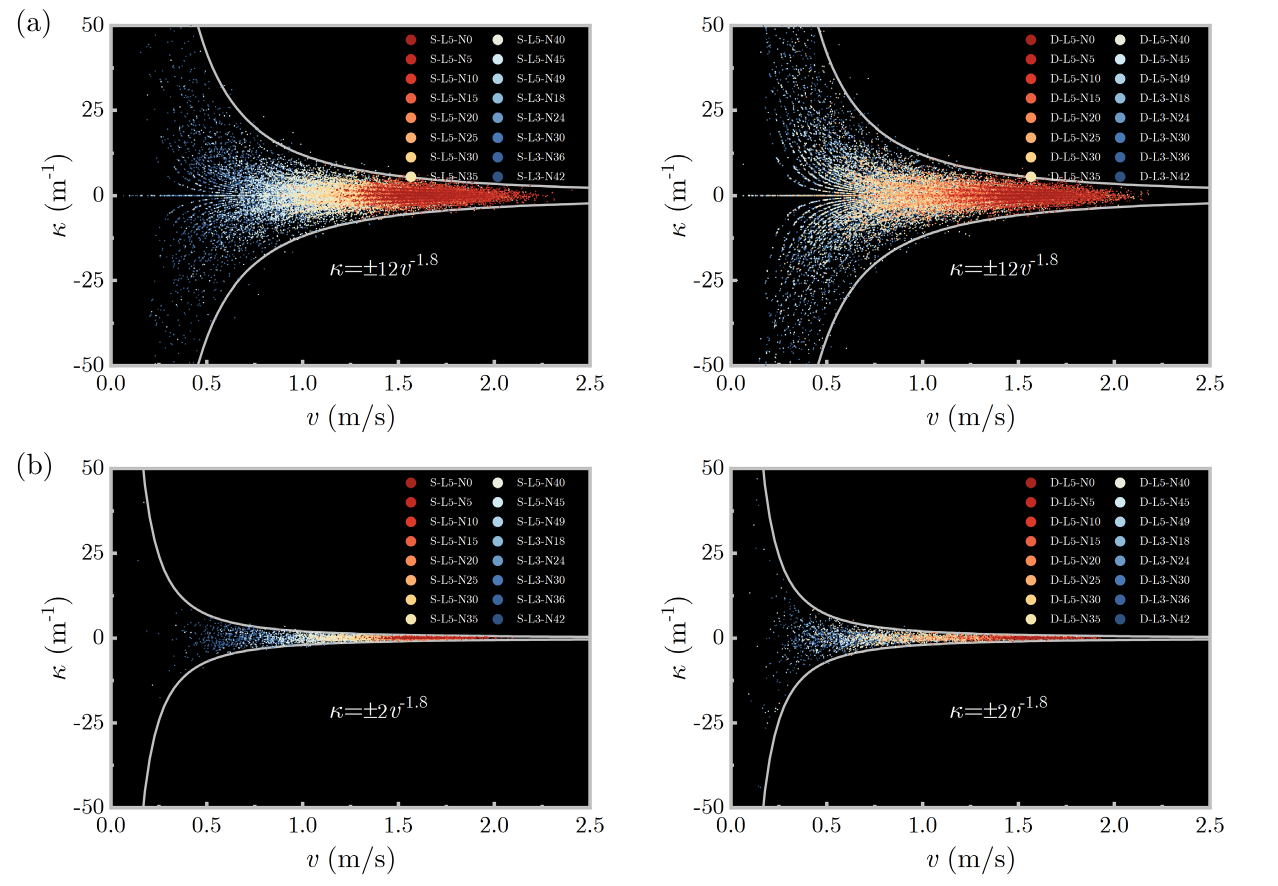}
\caption{Power-law scaling between critical curvature and speed: (a) frame-by-frame sampling, (b) sampling every 10 frames.}
\label{fig12}
\end{figure}

\begin{figure}[ht!]
\centering
\includegraphics[scale=0.66]{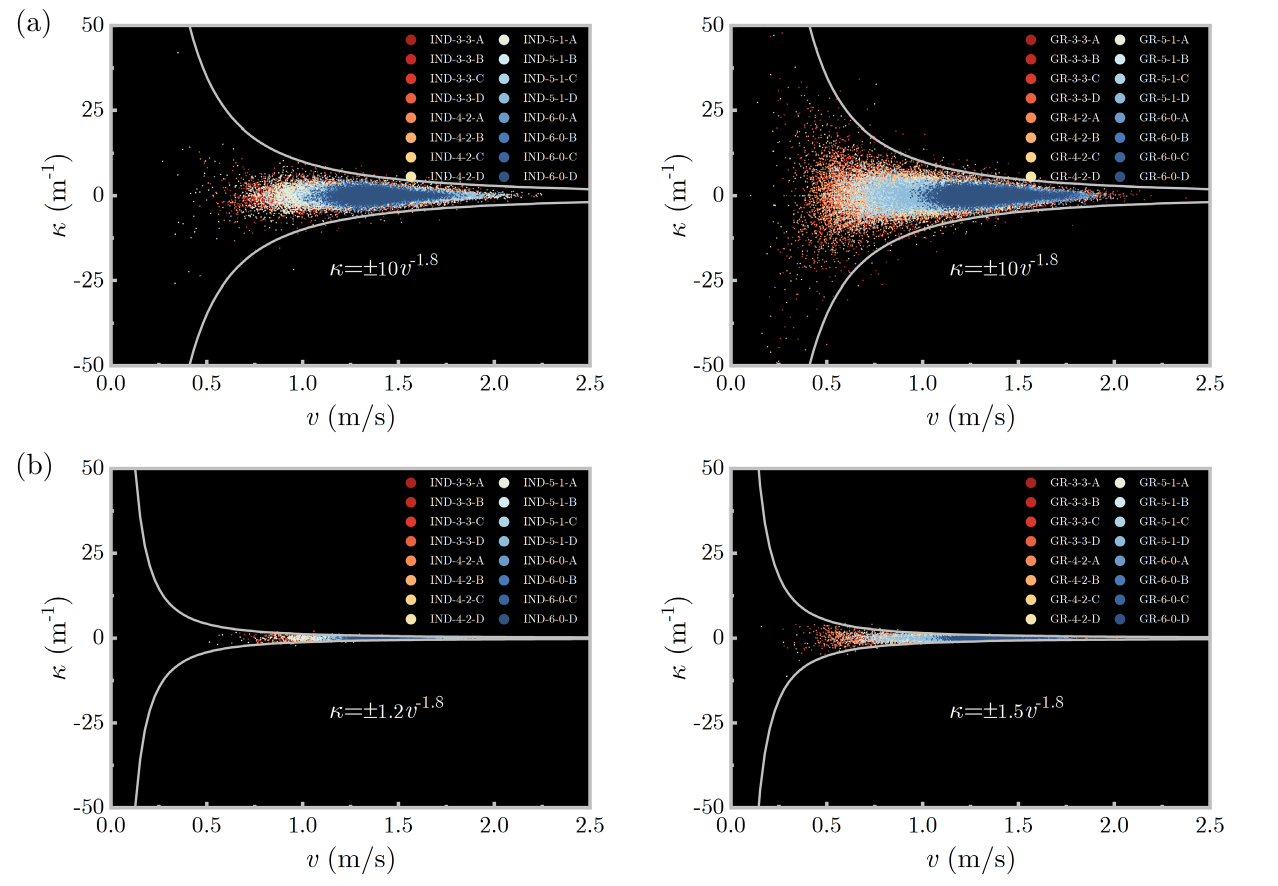}
\caption{Power-law scaling between critical curvature and speed: (a) frame-by-frame sampling, (b) sampling every 10 frames.}
\label{fig13}
\end{figure}

\begin{figure}[ht!]
\centering
\includegraphics[scale=0.66]{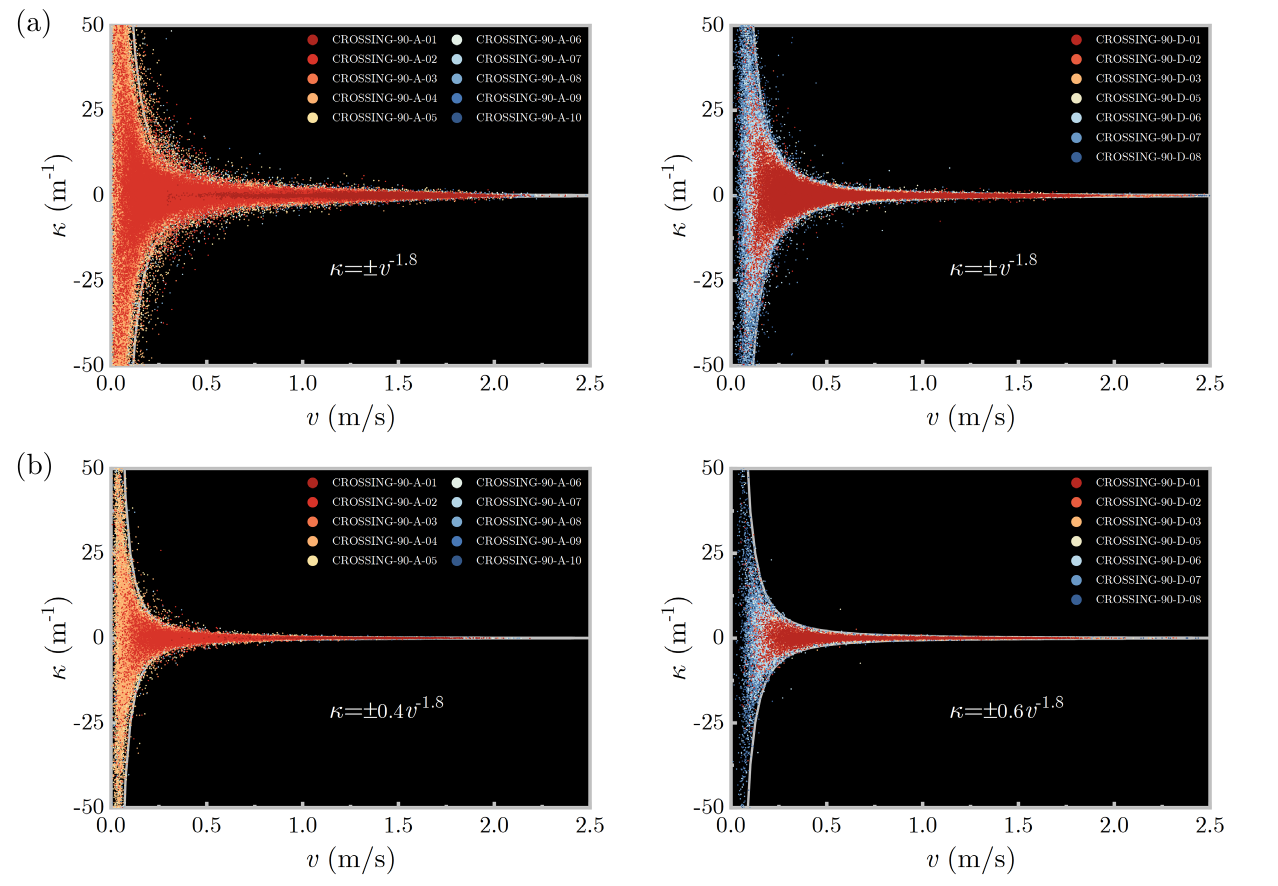}
\caption{Power-law scaling between critical curvature and speed: (a) frame-by-frame sampling, (b) sampling every 10 frames.}
\label{fig14}
\end{figure}

\begin{figure}[ht!]
\centering
\includegraphics[scale=0.66]{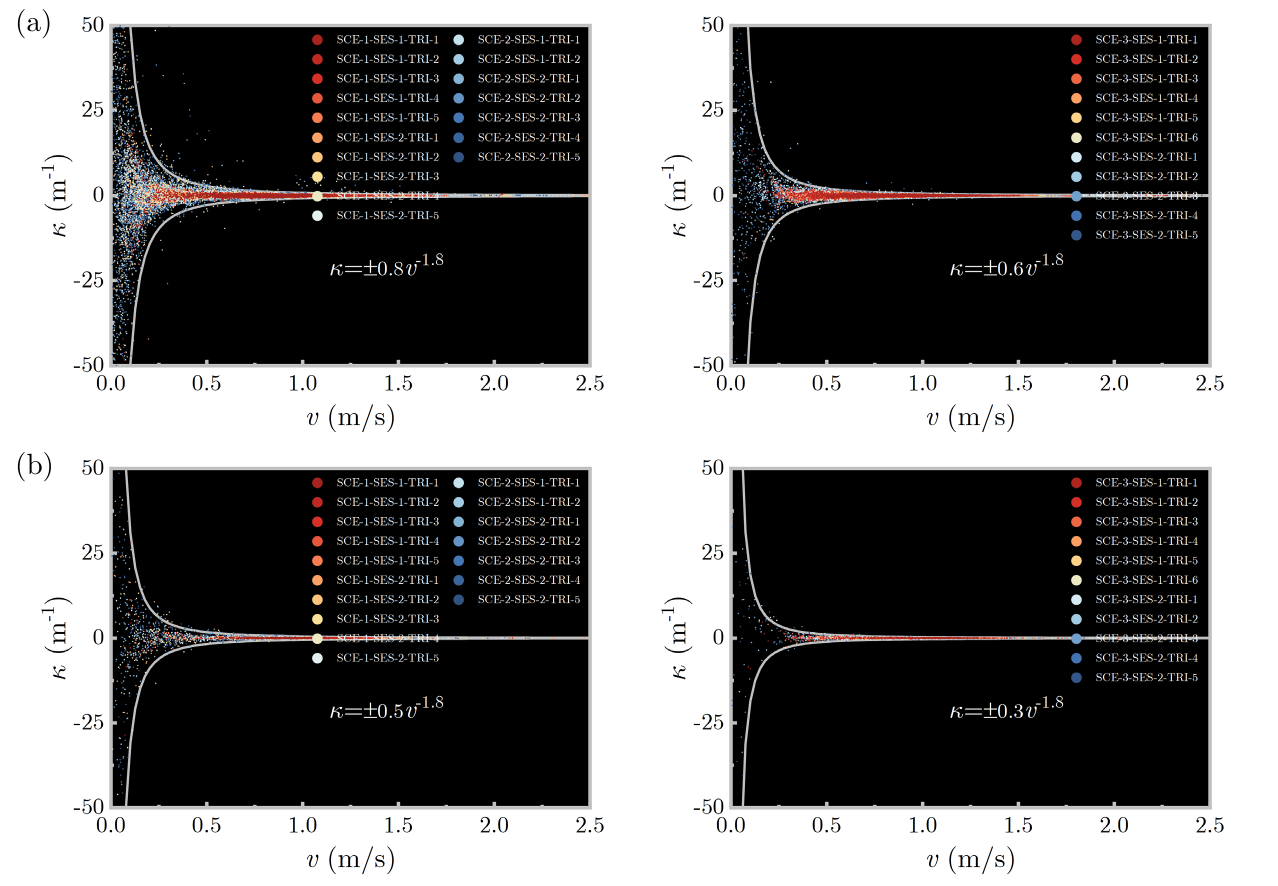}
\caption{Power-law scaling between critical curvature and speed: (a) frame-by-frame sampling, (b) sampling every 10 frames.}
\label{fig15}
\end{figure}

\begin{figure}[ht!]
\centering
\includegraphics[scale=0.66]{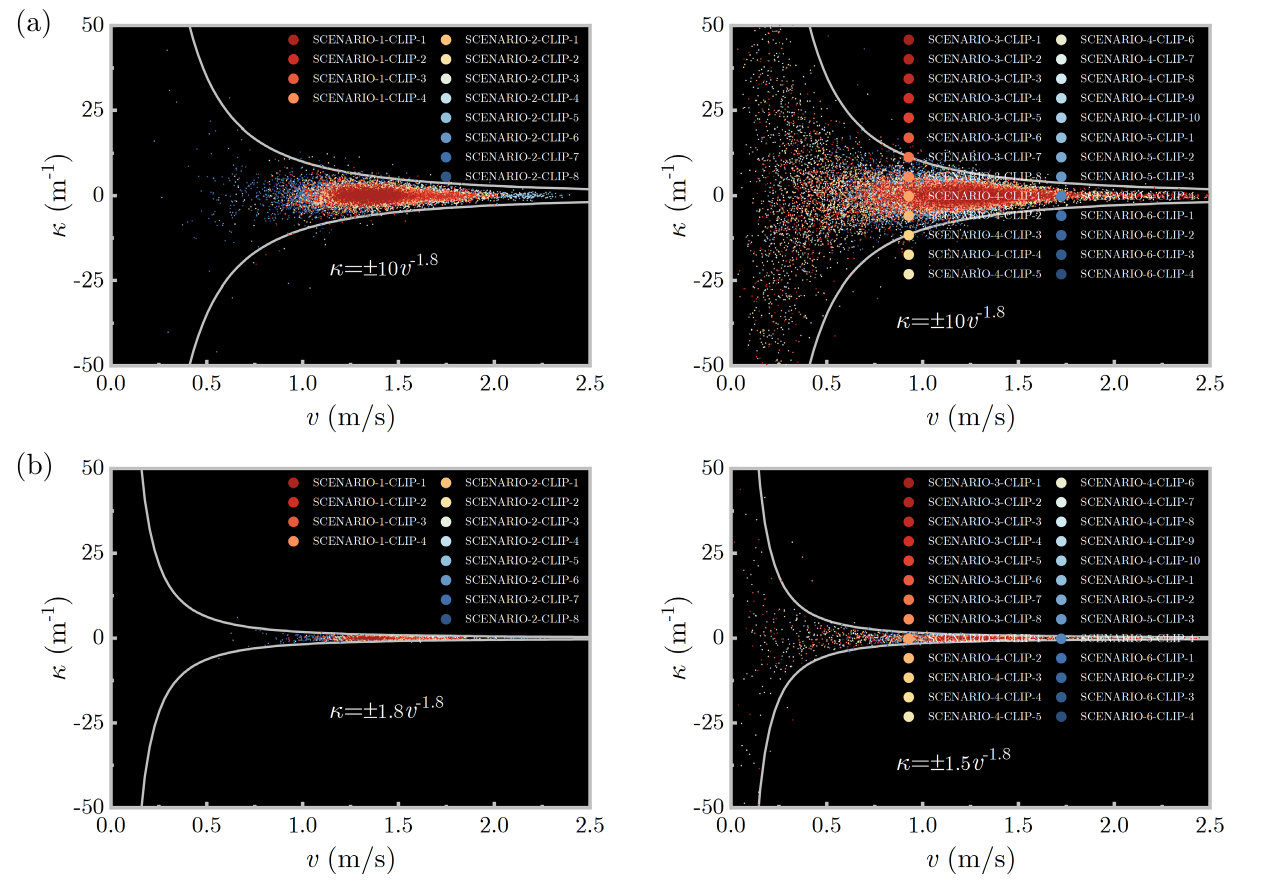}
\caption{Power-law scaling between critical curvature and speed: (a) frame-by-frame sampling, (b) sampling every 10 frames.}
\label{fig16}
\end{figure}

\begin{figure}[ht!]
\centering
\includegraphics[scale=0.66]{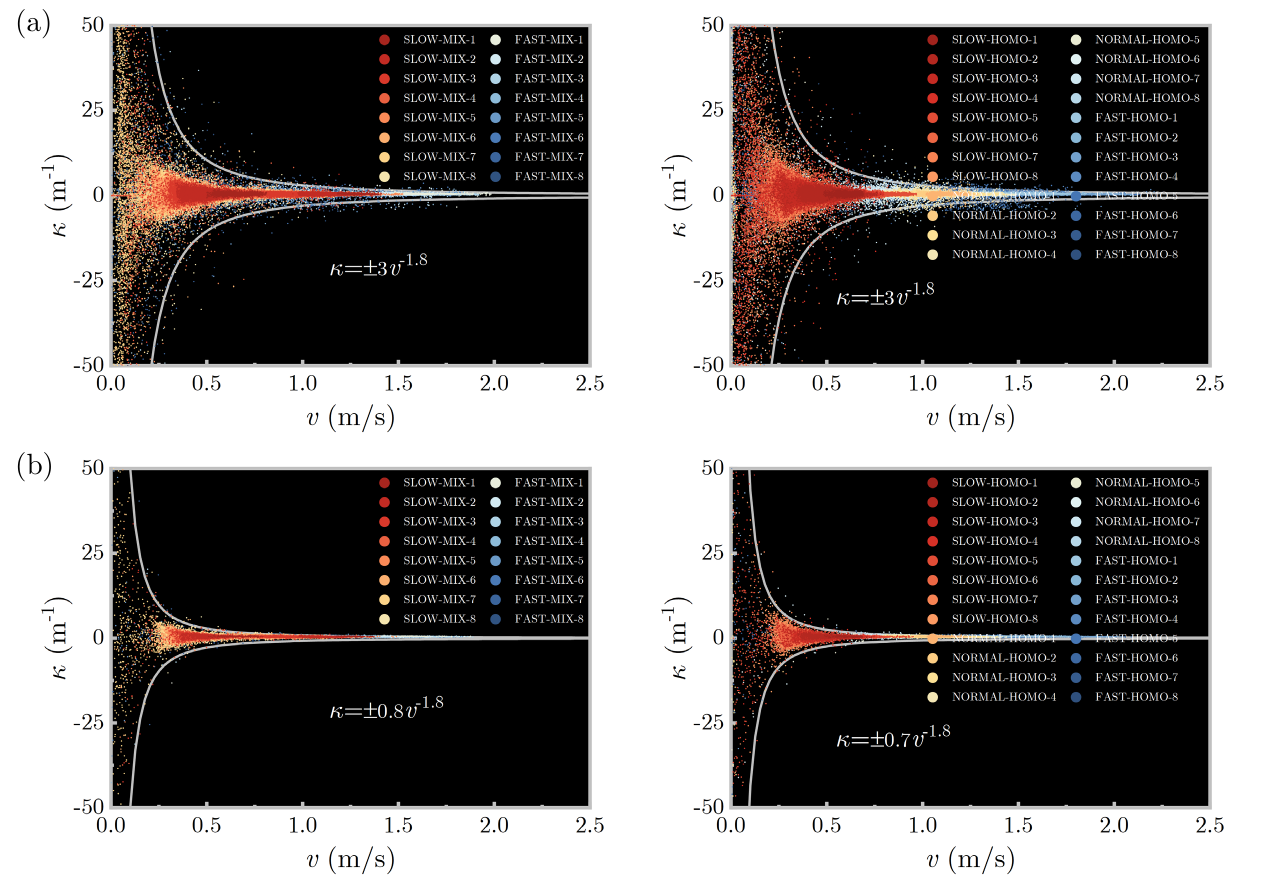}
\caption{Power-law scaling between critical curvature and speed: (a) frame-by-frame sampling, (b) sampling every 10 frames.}
\label{fig17}
\end{figure}

\begin{figure}[ht!]
\centering
\includegraphics[scale=0.66]{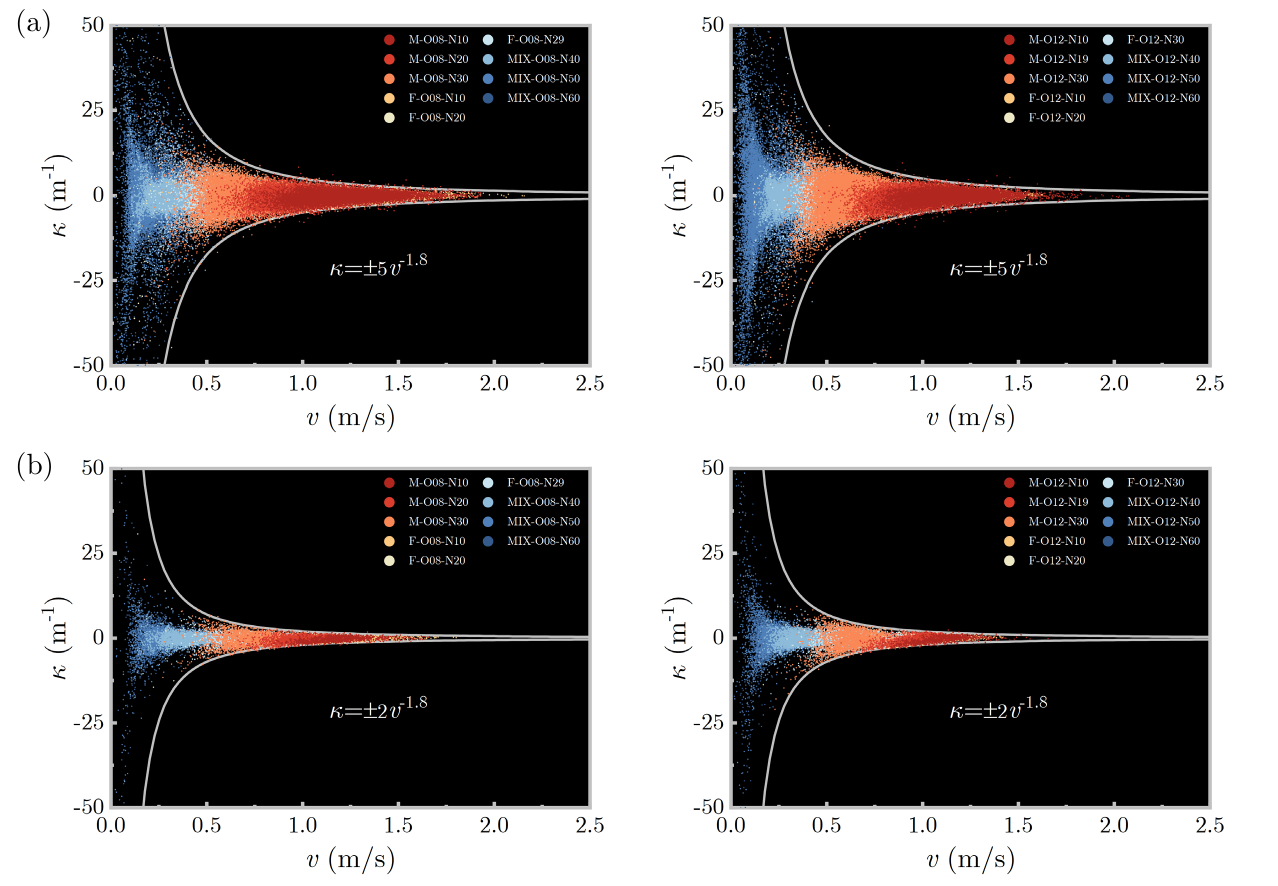}
\caption{Power-law scaling between critical curvature and speed: (a) frame-by-frame sampling, (b) sampling every 10 frames.}
\label{fig18}
\end{figure}

\begin{figure}[ht!]
\centering
\includegraphics[scale=0.66]{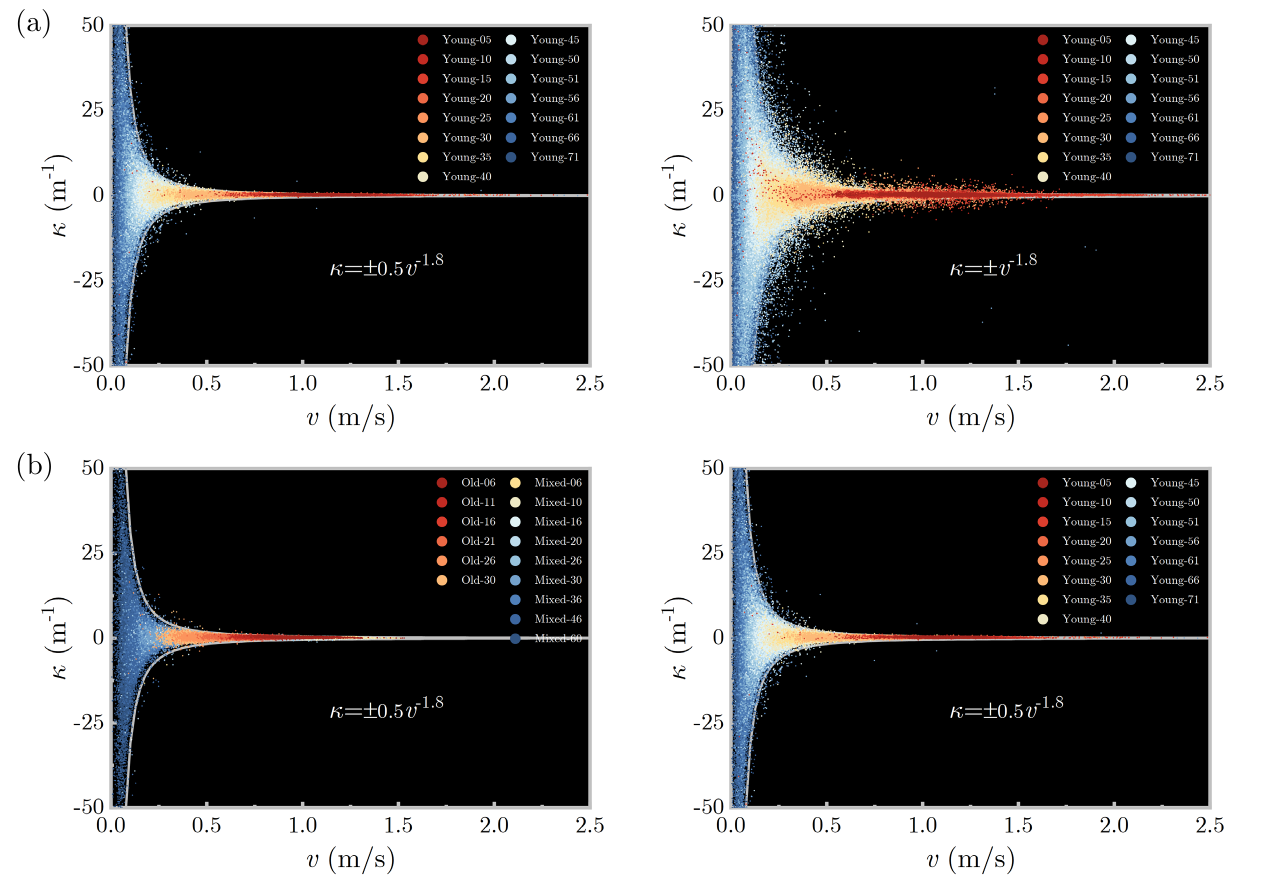}
\caption{Power-law scaling between critical curvature and speed: (a) frame-by-frame sampling, (b) sampling every 10 frames.}
\label{fig19}
\end{figure}

\begin{figure}[ht!]
\centering
\includegraphics[scale=0.66]{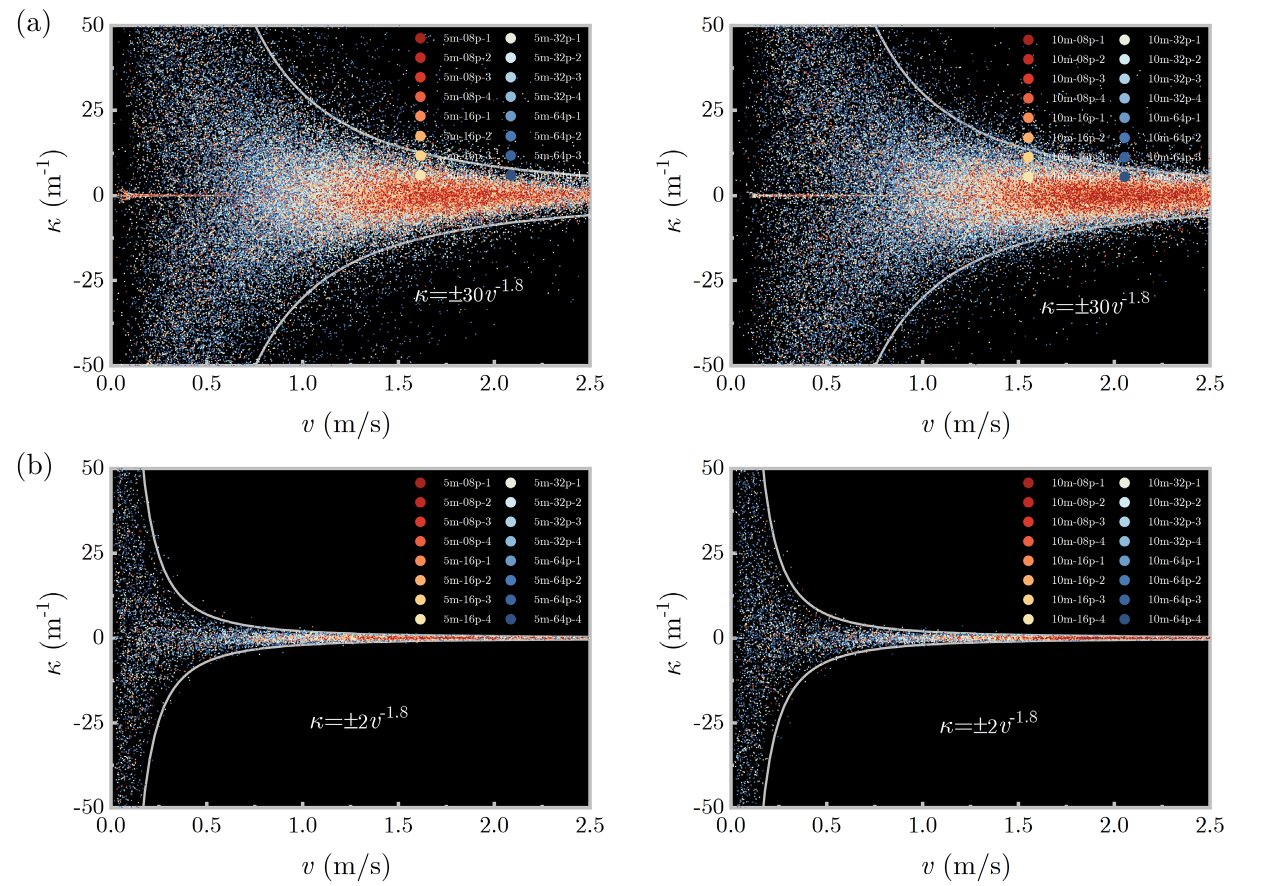}
\caption{Power-law scaling between critical curvature and speed: (a) frame-by-frame sampling, (b) sampling every 10 frames.}
\label{fig20}
\end{figure}

In this section, we present the statistical results for each experimental group.  The corresponding statistical results for frame-by-frame sampling and 10-frame sampling are displayed in subplots (a) and (b) in every scatter diagrams (Figs. \ref{fig12} - \ref{fig20}), respectively.

For each figure, a subjectively defined trend curve and its corresponding function were established. The scatter data distribution exhibits a clear and consistent trend that closely follows the defined power-law curve with a fixed exponent of -1.8. This indicates a decay relationship between the maximum curvature of pedestrian trajectories (regarded as a critical value) and speed. Also, the results imply that as pedestrians speed up, corresponding maximum angular velocity decreases according to a power-law function with an exponent of -0.8, as presented in Eq. \ref{3}.

\begin{equation}\label{3}
\kappa_{\text{crit}} \propto \pm v^{-1.8} \triangleq  \omega_{\text{crit}} \propto \pm v^{-0.8}
\end{equation}

For detailed statistical results of the curves in each experiment, refer to Figs. \ref{fig12}–\ref{fig20}. Under the sampling configurations of frame-by-frame sampling and sampling every 10 frames, the scale effect results in a linear expansion or contraction of the trend lines associated with the scatter data, manifested as variations in the constant coefficients of the trend curves.

\section{Discussion} \label{section4}

To investigate the relationship between speed and direction during pedestrian motion, we conducted a statistical analysis on tens of thousands of trajectories from nine datasets. To minimize potential errors caused by data noise or statistical artifacts, the trajectory data were first subjected to various preprocessing methods, including mean filtering, adaptive filtering, Kalman filtering, as well as analysis of raw data for comparison. Furthermore, to avoid biases introduced by sampling frequency, we performed independent statistics at different sampling intervals: every frame and every ten frames. Despite these variations in preprocessing and sampling, the results consistently revealed a universal pattern: a power-law scaling relationship between the critical angular velocity and speed.

The statistical data were obtained through direct sampling of trajectory data, specifically focusing on changes in position and orientation. This constrained relationship presents a basic property of pedestrian motion. Overall, our results demonstrate that pedestrian speed is associated with a strictly bounded range of angular velocities governing pedestrian motion, and this range exhibits a power-law decay as speed increases.

From a broader perspective, these findings suggest that both the accuracy of trajectory prediction and the complexity of pedestrian motion are intrinsically linked to walking speed. Our results provide an explanation for why linear regression methods often perform well in trajectory prediction tasks under normal conditions, given their simplicity \citep{gupta2018social}. In contrast, at lower speeds, pedestrian movement tends to become more complex, often characterized by frequent changes in direction. Due to the complexity of pedestrian interactions in high-density crowds, accurately predicting trajectories remains a formidable challenge.

\centerline{}
\section*{Data Availability}
Data can be found at: \url{https://drive.google.com/drive/folders/1NYVnRp0z8VPuskfezMr51gB-sraOf6Iq?usp=drive_link} (Google Drive).

\centerline{}
\section*{Acknowledgments}
This work was supported by the National Natural Science Foundation of China (Grant No. 52072286, 71871189, 51604204), and the Fundamental Research Funds for the Central Universities (Grant No. 2022IVA108).

\bibliographystyle{aasjournal}

\begin{thebibliography}{}
\expandafter\ifx\csname natexlab\endcsname\relax\def\natexlab#1{#1}\fi
\providecommand{\url}[1]{\href{#1}{#1}}
\providecommand{\dodoi}[1]{doi:~\href{http://doi.org/#1}{\nolinkurl{#1}}}
\providecommand{\doeprint}[1]{\href{http://ascl.net/#1}{\nolinkurl{http://ascl.net/#1}}}
\providecommand{\doarXiv}[1]{\href{https://arxiv.org/abs/#1}{\nolinkurl{https://arxiv.org/abs/#1}}}

\bibitem[{Bacik {et~al.}(2023)Bacik, Bacik, \& Rogers}]{bacik2023lane}
Bacik, K.~A., Bacik, B.~S., \& Rogers, T. 2023, Science, 379, 923, \dodoi{10.1126/science.add8091}

\bibitem[{Cao {et~al.}(2017)Cao, Seyfried, Zhang, Holl, \& Song}]{cao2017fundamental}
Cao, S., Seyfried, A., Zhang, J., Holl, S., \& Song, W. 2017, Journal of Statistical Mechanics: Theory and Experiment, 2017, 033404, \dodoi{10.1088/1742-5468/aa620d}

\bibitem[{Cao {et~al.}(2016)Cao, Zhang, Salden, Ma, Shi, \& Zhang}]{cao2016pedestrian}
Cao, S., Zhang, J., Salden, D., {et~al.} 2016, Physical Review E, 94, 012312, \dodoi{10.1103/PhysRevE.94.012312}

\bibitem[{Cordes {et~al.}(2023)Cordes, Chraibi, \& Tordeux}]{cordes2023single}
Cordes, J., Chraibi, M., \& Tordeux, A. 2023, Crowd Dynamics, Volume 4: Analytics and Human Factors in Crowd Modeling, 143, \dodoi{10.1007/978-3-031-46359-4_6}

\bibitem[{Dias {et~al.}(2013)Dias, Sarvi, Shiwakoti, \& Ejtemai}]{dias2013experimental}
Dias, C., Sarvi, M., Shiwakoti, N., \& Ejtemai, O. 2013, in Proceedings of 36th Australasian Transport Research Forum, 1--11, \dodoi{10.1038/s41598-018-36711-7}

\bibitem[{Dos’~Santos {et~al.}(2018)Dos’~Santos, Thomas, Comfort, \& Jones}]{dos2018effect}
Dos’~Santos, T., Thomas, C., Comfort, P., \& Jones, P.~A. 2018, Sports medicine, 48, 2235, \dodoi{10.1007/s40279-018-0968-3}

\bibitem[{Elliott {et~al.}(2012)Elliott, Simms, \& Wood}]{elliott2012pedestrian}
Elliott, J., Simms, C.~K., \& Wood, D.~P. 2012, Accident Analysis \& Prevention, 45, 342, \dodoi{10.1016/j.aap.2011.07.022}

\bibitem[{Farina {et~al.}(2017)Farina, Fontanelli, Garulli, Giannitrapani, \& Prattichizzo}]{farina2017walking}
Farina, F., Fontanelli, D., Garulli, A., Giannitrapani, A., \& Prattichizzo, D. 2017, PloS one, 12, e0169734, \dodoi{10.1371/journal.pone.0169734}

\bibitem[{Feliciani \& Nishinari(2016)}]{feliciani2016empirical}
Feliciani, C., \& Nishinari, K. 2016, Physical Review E, 94, 032304, \dodoi{10.1103/PhysRevE.94.032304}

\bibitem[{Fujita {et~al.}(2019)Fujita, Feliciani, Yanagisawa, \& Nishinari}]{fujita2019traffic}
Fujita, A., Feliciani, C., Yanagisawa, D., \& Nishinari, K. 2019, Physical Review E, 99, 062307, \dodoi{10.1103/PhysRevE.99.062307}

\bibitem[{Greenshields {et~al.}(1935)Greenshields, Bibbins, Channing, \& Miller}]{greenshields}
Greenshields, B.~D., Bibbins, J.~R., Channing, W.~S., \& Miller, H.~H. 1935, in Highway Research Board Proceedings, Vol.~14, {Washington, DC}, 448--477

\bibitem[{Gupta {et~al.}(2018)Gupta, Johnson, Fei-Fei, Savarese, \& Alahi}]{gupta2018social}
Gupta, A., Johnson, J., Fei-Fei, L., Savarese, S., \& Alahi, A. 2018, in Proceedings of the IEEE conference on computer vision and pattern recognition, 2255--2264

\bibitem[{Hicheur {et~al.}(2005)Hicheur, Vieilledent, Richardson, Flash, \& Berthoz}]{hicheur2005velocity}
Hicheur, H., Vieilledent, S., Richardson, M.~J., Flash, T., \& Berthoz, A. 2005, Experimental brain research, 162, 145, \dodoi{10.1007/s00221-004-2122-8}

\bibitem[{Hoogendoorn {et~al.}(2011)Hoogendoorn, Campanella, \& Daamen}]{hoogendoorn2011fundamental}
Hoogendoorn, S., Campanella, M., \& Daamen, W. 2011, in Pedestrian and Evacuation Dynamics (Springer), 255--264, \dodoi{10.1007/978-1-4419-9725-8_23}

\bibitem[{Huber {et~al.}(2014)Huber, Su, Kr{\"u}ger, Faschian, Glasauer, \& Hermsd{\"o}rfer}]{huber2014adjustments}
Huber, M., Su, Y.-H., Kr{\"u}ger, M., {et~al.} 2014, PloS one, 9, e89589, \dodoi{10.1371/journal.pone.0089589}

\bibitem[{Kleinmeier {et~al.}(2020)Kleinmeier, K{\"o}ster, \& Drury}]{kleinmeier2020agent}
Kleinmeier, B., K{\"o}ster, G., \& Drury, J. 2020, Journal of the Royal Society Interface, 17, 20200396, \dodoi{10.1098/rsif.2020.0396}

\bibitem[{MathWorks(2023)}]{mathworks_lms_nlms}
MathWorks. 2023, Enhance a Signal Using {LMS} and Normalized {LMS} Algorithms.
\newblock \url{https://www.mathworks.com/help/dsp/ug/enhance-a-signal-using-lms-and-normalized-lms-algorithms.html}

\bibitem[{Nicolas {et~al.}(2019)Nicolas, Kuperman, Iba{\~n}ez, Bouzat, \& Appert-Rolland}]{nicolas2019mechanical}
Nicolas, A., Kuperman, M., Iba{\~n}ez, S., Bouzat, S., \& Appert-Rolland, C. 2019, Scientific reports, 9, 105, \dodoi{10.1038/s41598-018-36711-7}

\bibitem[{Parisi {et~al.}(2016)Parisi, Negri, \& Bruno}]{parisi2016experimental}
Parisi, D.~R., Negri, P.~A., \& Bruno, L. 2016, Physical Review E, 94, 022318, \dodoi{10.1103/PhysRevE.94.022318}

\bibitem[{Saberi \& Mahmassani(2014)}]{saberi2014exploring}
Saberi, M., \& Mahmassani, H.~S. 2014, Transportation research record, 2421, 31, \dodoi{10.3141/2421-04}

\bibitem[{Seyfried {et~al.}(2005)Seyfried, Steffen, Klingsch, \& Boltes}]{seyfried2005fundamental}
Seyfried, A., Steffen, B., Klingsch, W., \& Boltes, M. 2005, Journal of Statistical Mechanics: Theory and Experiment, 2005, P10002, \dodoi{10.1088/1742-5468/2005/10/P10002}

\bibitem[{Wang {et~al.}(2023)Wang, Lv, Jiang, Fang, \& Ma}]{wang2023exploring}
Wang, J., Lv, W., Jiang, H., Fang, Z., \& Ma, J. 2023, arXiv preprint arXiv:2311.04827, \dodoi{10.48550/arXiv.2311.04827}

\bibitem[{Xiao {et~al.}(2019)Xiao, Gao, Jiang, Li, Qu, \& Huang}]{xiao2019investigation}
Xiao, Y., Gao, Z., Jiang, R., {et~al.} 2019, Transportation Research Part C: Emerging Technologies, 103, 174, \dodoi{10.1016/j.trc.2019.04.007}

\bibitem[{Yang {et~al.}(2019)Yang, Li, Redmill, \& {\"O}zg{\"u}ner}]{yang2019top}
Yang, D., Li, L., Redmill, K., \& {\"O}zg{\"u}ner, {\"U}. 2019, in 2019 IEEE Intelligent Vehicles Symposium (IV), IEEE, 899--904, \dodoi{10.1109/IVS.2019.8814092}

\bibitem[{Zhang {et~al.}(2012)Zhang, Klingsch, Schadschneider, \& Seyfried}]{zhang2012ordering}
Zhang, J., Klingsch, W., Schadschneider, A., \& Seyfried, A. 2012, Journal of Statistical Mechanics: Theory and Experiment, 2012, P02002, \dodoi{10.1088/1742-5468/2012/00/P00000}

\end{thebibliography}

\end{document}